\documentclass[runningheads]{llncs}
\usepackage[T1]{fontenc}
\usepackage{newtxtext}       %
\usepackage[varvw]{newtxmath}       % selects Times Roman as basic font
\usepackage{graphicx}
\usepackage{comment}

\usepackage{tikz}
\usetikzlibrary{arrows.meta,calc,decorations.pathreplacing}
\usepackage{booktabs}
\newcommand{\OPT}{\mathrm{OPT}}
\newcommand{\E}{\mathbb{E}}
\newcommand{\Prob}{\mathbb{P}}
\newcommand{\one}{\mathbf{1}}
\newcommand{\R}{\mathbb{R}}
\DeclareMathOperator{\odd}{odd}
\newcommand{\rhoMP}{\rho_{\mathrm{MP}}}
\newcommand{\rhoOT}{\rho_{\mathrm{OT}}}
\newcommand{\Adbl}{\mathrm{ALG}_{\mathrm{dbl}}}
\newcommand{\Ajoin}{\mathrm{ALG}_{\mathrm{join}}}
\newcommand{\Alam}{\mathrm{ALG}_{\lambda}}
\newcommand{\Tr}{\mathcal{T}}
\newcommand{\Uc}{\mathcal{U}}
\newcommand{\Ac}{\mathcal{A}}
\newcommand{\LPrel}{\textup{(LP)}}
\newcommand{\dout}{\delta^{+}}
\newcommand{\din}{\delta^{-}}
\newcommand{\optlp}{\mathrm{opt}_{\mathrm{LP}}}
\usepackage{xcolor}
\usepackage{hyperref}
\usepackage{cleveref}

\hypersetup{pdftitle={Improved Approximation Algorithms for Multi-Path TSP and Ordered TSP},pdfauthor={Morteza Alimi and Michael Ruderer},
  colorlinks=true,linkcolor=blue!60!black,citecolor=green!40!black,urlcolor=blue!60!black}
\ifdefined\pdfsuppressptexinfo \pdfsuppressptexinfo=-1 \fi
\ifdefined\pdfinfoomitdate \pdfinfoomitdate=1 \fi
\ifdefined\pdftrailerid \pdftrailerid{} \fi
\newif\iffinal
\finaltrue % in the build that is uploaded

\newenvironment{restate}[2][restated]%
  {\begin{trivlist}\item[\hskip\labelsep\bfseries #2 (#1).]\itshape}%
  {\end{trivlist}}
\begin{document}
\title{Improved Approximation Algorithms\texorpdfstring{\\}{ } for Multi-Path TSP and Ordered TSP}
\titlerunning{Approximating Multi-Path TSP and Ordered TSP}
\author{Morteza Alimi \and Michael Ruderer}
\authorrunning{M. Alimi and M. Ruderer}
\institute{Institute of Computer Science, University of Augsburg, Germany\\
\email{morteza.alimi@uni-a.de, michael.ruderer@uni-a.de}}
\maketitle

\begin{abstract}
In Multi-Path TSP, given a metric and $k$ pairs of vertices $(s_i,t_i)$, where $k$ is
part of the input, we seek $s_i$--$t_i$ paths of minimum total length that together
visit all vertices. Ratios below $2$ were known only for graphic metrics and special
cases such as multiple depots ($s_i=t_i$); in general metrics, the best ratio was
$1+2e^{-1/2}\approx2.213$, by a randomized algorithm of B\"ohm, Friggstad, M\"omke and
Spoerhase [SODA 2025].
We give a deterministic algorithm with ratio
$\frac{5e-2}{3e-2}<1.8833$ relative to a restriction of their LP relaxation, bounding
its integrality gap by the same constant. For Ordered TSP, where a tour must visit $k$ given
vertices in a prescribed cyclic order, we compute tours within
$1+\frac{e}{2(e-1)}<1.7910$ times the value of the LP relaxation of Armbruster, Mnich
and N\"agele [APPROX 2024], improving $\frac32+\frac1e\approx1.868$ on both the ratio and the
integrality gap.

Our algorithms sample a tree for every pair from an LP solution, but keep only its
path between the two endpoints (for Ordered TSP, the pairs are consecutive given
vertices). The main challenge is to reconnect the vertices missed by these paths
without losing the savings from discarding the branches. We prove a \emph{threshold
connection bound}: for arbitrary nonnegative edge weights, a cheapest forest
connecting all vertices to the paths has expected weight at most $1/e$ times that of
a cheapest forest rooted at the endpoints plus $(1-1/e)$ times the expected weight of
the discarded branches. The proof uses the components formed by cheap edges to charge
the additional cost of missed coverage to discarded branches.
For Multi-Path TSP, we apply this bound in the weights of the two-speed primal--dual
growth process that Zhao, Liu and Xiao [arXiv 2026] designed for multiple depots, in
which the connecting forest also pays for part of correcting its degree parities.
Our \emph{reweighting theorem} shows that the dual slack of this process pays for the
reweighting. Finally, we use the LP's fractional
assignment of vertices to pairs to
repair deficient parity cuts. In particular, we extend their parity analysis from optimal
    solutions to fractional LP solutions, yielding an LP-relative guarantee for
Multi-Path TSP.
\end{abstract}

\section{Introduction}\label{sec:intro}

In \emph{Multi-Path TSP}, we are given a finite metric $(V,c)$, in which distinct
vertices may be at distance zero, and $k$ pairs of vertices $(s_1,t_1)$, \dots,
$(s_k,t_k)$, and we seek, for every $i$, a path from $s_i$ to $t_i$ such that the
paths together visit every vertex and their total cost is minimum; the number $k$ of
pairs is part of the input. The problem is also known as $k$-Person TSP
Path~\cite{BFMS}. 
It contains $s$--$t$ Path TSP ($k=1$) and the multiple-depot traveling
salesman problem (each pair consists of two copies of one
depot)~\cite{ADMPV,BFMS}.
Moreover, instances with bounded $k$ arise as intermediate problems in
algorithms for $s$--$t$ Path TSP~\cite{ADMPV,TV}.
In \emph{Ordered
TSP}, we are given $(V,c)$ and distinct vertices $d_1,\dots,d_k$, and we seek a
cheapest Hamiltonian cycle that visits $d_1,\dots,d_k$ in the given order. By
shortcutting, it is equivalent to Multi-Path TSP with the pairs $(d_1,d_2)$, \dots,
$(d_{k-1},d_k)$, $(d_k,d_1)$, and it is a special case of precedence-constrained TSP~\cite{AMN,BFMS}. We call the endpoints of the pairs $(s_i, t_i)$, and the vertices $d_i$,
\emph{terminals}.

For Ordered TSP, the ratios $\frac52$ of B\"ockenhauer, Hromkovi\v{c}, Kneis and
Kupke~\cite{BHKK}, obtained with the algorithm of Christofides and
Serdyukov~\cite{Christofides,Serdyukov}, and $\frac52-\frac2k$ of B\"ockenhauer, M\"omke
and Steinov\'a~\cite{BMS} were improved by Armbruster, Mnich and
N\"agele~\cite{AMN} and by B\"ohm, Friggstad, M\"omke and Spoerhase~\cite{BFMS} to
$\frac32+\frac1e\approx1.868$; the analysis of~\cite{AMN} is relative to its LP relaxation.
For Multi-Path TSP, \cite{BFMS} gave an LP-based algorithm with expected ratio
$1+2e^{-1/2}\approx2.213$, improving the folklore ratio~$3$ (double a cheapest forest
in which every component contains one terminal, and add the edges $s_it_i$). In
graphic metrics, the shortest-path metrics of connected unweighted graphs, Alimi, Dahlmeier, M\"omke, Pabst and Vargas Koch~\cite{ADMPV}
obtained ratio $2$ for Multi-Path TSP and $1+\frac{e}{2(e-1)}\approx1.791$ for
Ordered TSP, and Dahlmeier, Luqu\'e, Pabst and Vargas Koch~\cite{DLPV} recently
improved the graphic Multi-Path ratio to $\frac{11e-10}{6e-6}\approx1.9303$. In
general metrics, the ratio of~\cite{BFMS} was the best known for Multi-Path
TSP~\cite{DLPV}; ratios below $2$ were known only for special cases, such as one
pair~\cite{Zenklusen} and multiple depots~\cite{DKM,XR,ZLX}. For multiple depots, Xu and
Rodrigues~\cite{XR} gave ratio $\frac32$ for a fixed number of depots, Deppert, Kaul and
Mnich~\cite{DKM} a randomized $(\frac32+\varepsilon)$-approximation in time exponential in
the number of depots, and~\cite{DLPV} ratio $\frac32$ for graphic metrics. For the
prize-collecting versions, Alimi, M\"omke and Ruderer~\cite{AMR} gave ratios $2.097$
(Ordered) and $2.41$ (Multi-Path).

\subsection{Our results}

Let $\OPT$ denote the minimum cost of a feasible solution, and let $\optlp$ denote
the optimum value of the relaxation \LPrel{} defined in Section~\ref{sec:prelim}.

\begin{theorem}\label{thm:mp}
There is a deterministic polynomial-time algorithm for Multi-Path TSP
that computes a solution of cost at most $\rhoMP\cdot\optlp$, where \mbox{$\rhoMP=\frac{5e-2}{3e-2}<1.8833$}.
\end{theorem}

\noindent Theorem~\ref{thm:mp} improves the previous approximation ratio of
$1+2e^{-1/2}\approx 2.213$ for general metrics~\cite[Theorem~1]{BFMS} as well as the ratio
$\frac{11e-10}{6e-6}\approx 1.930$ for graphic metrics~\cite[Theorem~3]{DLPV}.
We use the primal--dual growth process that Zhao, Liu and Xiao~\cite{ZLX}
introduced for multiple-depot TSP, where they obtained an approximation
ratio of $1+1/\sqrt{2}\approx 1.707$.
%The analysis of~\cite{BFMS} is relative to an LP relaxation. 
Our relaxation \LPrel{}
uses the stroll formulation of~\cite{AMN} for Ordered TSP and is equivalent to a restriction of the
relaxation of~\cite{BFMS} (Appendix~\ref{app:lp}).
Theorem~\ref{thm:mp} is relative to \LPrel{} and improves
its integrality-gap bound from $1+2e^{-1/2}$; in graphic metrics, the bound~$2$
was known for essentially the same relaxation~\cite{ADMPV}.
%A fractional parity certificate uses the commodity coverage vectors to repair cuts on which half of the LP solution is insufficient (Section~\ref{sec:growth}).

\begin{corollary}\label{cor:gap-mp}
For Multi-Path TSP, let $\Delta=\sum_{i=1}^k c(s_i,t_i)$. Then
\[
  \OPT \le \min\bigl\{\rhoMP\cdot\optlp,\,
    2\optlp-(1-\tfrac{2}{e})\Delta\bigr\}.
\]
In particular, the integrality gap of \LPrel{} is at most $\rhoMP<1.8833$.
\end{corollary}

For Ordered TSP, \cite[Theorem~1]{AMN} and~\cite[Theorem~2]{BFMS} give ratio
$\frac32+\frac1e$, also an integrality-gap bound in~\cite[Theorem~6]{AMN}.
The following theorem obtains in general metrics the constant proved for graphic
metrics in~\cite[Theorem~1.2]{ADMPV}.

\begin{theorem}\label{thm:ot}
There is a deterministic polynomial-time algorithm for Ordered TSP
that computes a tour of cost at most $\rhoOT\cdot\optlp$, where
\mbox{$\rhoOT=1+\frac{e}{2(e-1)}<1.7910$}.
\end{theorem}

\begin{corollary}\label{cor:gap}
For Ordered TSP with $k\ge2$ ordered vertices, the integrality gap
of \LPrel{} is at most $\rhoOT<1.7910$.
\end{corollary}

The graphic algorithm of~\cite{ADMPV} attaches each missed vertex
separately by a doubled edge. Our threshold connection bound
(Theorem~\ref{thm:threshold}) allows us to account for the cost of
attaching several missed vertices together, even in general metrics.

\subsection{Overview of the techniques}\label{sec:overview}

Both algorithms are variants of the algorithm of Christofides and
Serdyukov~\cite{Christofides,Serdyukov}. They choose a path between the endpoints of each
pair $(s_i,t_i)$, attach all remaining vertices to these paths by a forest, and then make
the degrees of this forest even by adding further edges. Finally, the resulting multigraph
can be shortcut into a feasible solution. The $s_i$--$t_i$ paths are sampled from an LP
solution.

Our algorithms use the same tree-sampling step as~\cite{AMN,BFMS}.
For each pair $(s_i,t_i)$, an optimal LP solution gives a distribution
over trees containing $s_i$ and $t_i$.
We independently sample one
tree $\Tr_i$ for each pair.
Every vertex belongs to at least one
sampled tree in expectation, and is missed by all
trees with probability at most $1/e$.
The expected cost of connecting the missed
vertices is then at most $1/e$ times the cost of a cheapest forest rooted at the terminal set
$T$; in~\cite{BFMS}, this is the Bridge Lemma (Lemma~\ref{lem:bridge}), which goes back to
Byrka et al.~\cite{BGRS}. 
The algorithms of~\cite{AMN,BFMS} include the sampled trees in the
solution, at an expected cost of up to the LP value.
For Ordered TSP, parity correction costs at most $\optlp/2$, as in
the LP-relative analysis of Christofides' algorithm~\cite{SW,Wolsey}.
Together with the cost of connecting the missed vertices, this gives
an expected total cost of at most $(1+\tfrac1e+\tfrac12)\optlp$.
The construction of~\cite{BFMS} duplicates the sampled-tree edges outside
the prescribed endpoint paths, allowing those edges to be traversed out and back.
These algorithms can all be derandomized by conditional expectations
(see Appendix~\ref{app:tools} for ours).

\paragraph{Keeping only the trunks.} Following the path-sampling approach
of~\cite{ADMPV} for graphic metrics, we keep, of each sampled tree $\Tr_i$, only its
path $Q_i$ between $s_i$ and $t_i$, which we call its \emph{trunk}.
We call a minimum-cost forest rooted at the trunk vertices a
\emph{connection forest} and use it to attach the remaining vertices
to the trunks.
Discarding the \emph{branches}
$R_i$ (the other edges of $\Tr_i$) saves their cost, but a vertex may now be missed
with probability close to one, so a bound that depends only on the probability with
which each vertex is missed, such as the Bridge Lemma, gives little. 
Our main lemma (Theorem~\ref{thm:threshold}) bounds the cost of
connecting the missed vertices by taking the weight of the discarded
branches into account.
Let $Q$ and $R$ be the multiset
unions of all trunks and of all branches. For arbitrary edge weights $w\ge0$, the cost
$f_w(Q)$ of a cheapest connection forest
satisfies
\begin{equation}\label{eq:intro-threshold}
 \E[f_w(Q)]\;\le\;\tfrac1e\,f_w(T)+\bigl(1-\tfrac1e\bigr)\,\E[w(R)],
\end{equation}
where $f_w(T)$ is the cost of a cheapest forest rooted at the terminals. The proof
writes $f_w(Q)$, as in Kruskal's algorithm~\cite{Kruskal}, as the integral over $t$ of
the number of components of the graph $\{e:w_e\le t\}$ that contain no terminal and
are missed by all trunks. A component $U$ is missed with probability at most
$e^{-\mu(U)}$, where $\mu(U)$ is the expected number of trunks that intersect $U$. If
$\mu(U)<1$, the coverage of the trees, oriented away from the $s_i$, forces an
expected number of at least $1-\mu(U)$ branch arcs that enter $U$, and a branch arc
$a$ enters a component of $\{e:w_e\le t\}$ only if $t<w_a$.

\paragraph{Ordered TSP.} The trunks of the pairs $(d_i,d_{i+1})$ form a closed walk
through $d_1,\dots,d_k$ in the prescribed order. The expected costs of the trunks and of
the branches add up to the LP value $\optlp$, and \eqref{eq:intro-threshold} with $w=c$ bounds the
expected connection cost by $\frac1e\optlp+(1-\frac1e)(\optlp-\E[c(Q)])$. 
For a fixed connection forest $K$, a cheapest join costs at most
$\min\{c(K),\frac12\,\optlp\}$, because $K$ itself is a feasible join on its odd vertices.
Thus the same completion has two expected cost bounds,
$2\,\optlp-(1-\frac2e)\E[c(Q)]$ and $\frac32\,\optlp+\frac1e\E[c(Q)]$.
The first decreases and the second increases with $\E[c(Q)]$; when they are equal,
both are $(1+\frac{e}{2(e-1)})\optlp$.

\paragraph{Multi-Path TSP.} The doubled-forest bound is strongest when $\Delta$ or
$\optlp-f_c(T)$ is large (Lemma~\ref{lem:doubling}). For the remaining regime, we use
the two-speed growth process of Zhao, Liu and Xiao~\cite{ZLX}, in the primal--dual
framework of Goemans and Williamson~\cite{GW}. For a parameter $\lambda\in[0,1)$,
it labels each vertex $v$ by the time $h(v)$ at which it connects to a terminal and defines
$\theta_{uv}=(c_{uv}-\min\{h(u),h(v)\})^+$ and $w=c+\lambda\theta$,
where $a^+=\max\{a,0\}$.
Section~\ref{sec:growth} gives the growth rules.

Our parity certificate uses the fractional coverage $y^i_v$ of vertex $v$ by pair $i$.
For each nonempty set $A$ of non-terminals, the packing--covering lemma bounds
the total weight of the selected edge fractions inside $A$ by
$\sum_i\max_{v\in A}y^i_v-1$.
By applying this bound at each time $t$
to the set of still unrooted vertices and integrating, we bound the label
cost of the repair by a fractional \emph{sum of peaks}
(Lemmas~\ref{lem:source} and~\ref{lem:parity}).
Combined with doubling, the resulting
bound charges a connection forest $K$ through $\frac2{1+\lambda}w(K)$.
Hence we optimize the forest in $w$, but compute joins and shortcutting in $c$.
Writing $X$ for the aggregate edge vector of the LP solution, our reweighting theorem gives
$w(X)\le\optlp+\frac{2\lambda}{1-\lambda}(\optlp-f_w(T))$:
the same forest value controls both the connection bound and the cost increase.

The algorithm compares this \emph{reweighted candidate} with the \emph{doubling
candidate}. At $\lambda=\frac2{15}$ their bounds yield $\frac{5e-2}{3e-2}$;
Appendix~\ref{app:constants} explains the parameter choice and the balance determining
the constant.

\paragraph{What is standard and what is new.} We apply the stroll relaxation and tree
decomposition of~\cite{AMN} (building on~\cite{BN} and the branching-packing result of~\cite{BFJ}) separately to the prescribed pairs.
Lemma~\ref{lem:bridge} is
the Bridge Lemma of~\cite{BFMS} (from~\cite{BGRS}; see also~\cite[Lemma~9]{AMN}). Sampling only paths between the endpoints is the
approach of~\cite{ADMPV} for graphic metrics. The growth process, labels, surcharge and reweighted cost are from~\cite{ZLX}.
Their parity argument~\cite[Lemma~3]{ZLX} uses an integral tour partition; our fractional coverage
certificate makes both the parity and peak bounds LP-relative
(Lemmas~\ref{lem:source} and~\ref{lem:parity}). Its packing--covering lemma and
full proofs are in Appendix~\ref{app:growth}. Theorems~\ref{thm:threshold}
and~\ref{thm:reweight} and their combination give the remaining improvements.

\section{Preliminaries}\label{sec:prelim}

Throughout the paper, let $G=(V,E)$ be a complete undirected graph
with nonnegative edge costs $c\in\mathbb{R}_{\ge0}^E$ satisfying
the triangle inequality. For the polynomial-time claims, costs are rational and encoded in binary.
For arbitrary $x\in\R^E$ and $E_0\subseteq E$, we write $x(E_0)=\sum_{e\in E_0}x_e$, counting multiplicity for multisets.
For $U\subseteq V$, let $\delta(U)$ be the edges with exactly one endpoint in $U$,
and write $\delta(v)=\delta(\{v\})$ and $a^+=\max\{a,0\}$.

\paragraph{The two problems in common form.} For Multi-Path TSP we assume, by
introducing colocated copies, that the $2k$ endpoints are distinct vertices. We
let $T=\{s_1,t_1,\dots,s_k,t_k\}$ be the set of \emph{terminals}, and $N=V\setminus T$ the set of \emph{non-terminals}. By shortcutting, some optimal solution consists of
paths $O_1,\dots,O_k$ such that $O_i$ goes from $s_i$ to $t_i$, contains no other
terminal, and every vertex of $N$ lies on exactly one of the paths.  
We fix such
a solution and write $\OPT=\sum_ic(O_i)$.
For Ordered TSP, let $d_1,\dots,d_k$ be the prescribed terminals in the given order.
We consider the pairs
$(s_i,t_i)=(d_i,d_{i+1})$ for $i\in[k]$, with $d_{k+1}=d_1$.\footnote{We may assume $k\ge3$. For
$k\le 2$, the problem is metric TSP.}
Thus, we have $T=\{d_1,\dots,d_k\}$, and $N=V\setminus T$.
In both problems,
$\OPT$ denotes the optimum value, and $\Delta=\sum_ic(s_i,t_i)$.

\paragraph{Forests and joins.} For nonempty $A\subseteq V$ and weights
$w\in\R^E_{\ge0}$, an \emph{$A$-rooted forest} is a forest with vertex set $V$ in
which every component contains exactly one vertex of $A$. We denote the
minimum $w$-cost of an $A$-rooted forest by $f_w(A)$. We call the vertices of $A$
the \emph{roots}. 
Identifying all roots in $G$ gives the multigraph $G/A$.
The $A$-rooted forests of $G$
correspond to the spanning trees of $G/A$.
%After identifying all vertices of $A$, these forests are the spanning trees, so a cheapest one can be computed inpolynomial time.
For a subgraph (or multigraph) $P$ of $G$, such as a path or a union of paths, we write
$f_w(P):=f_w(V(P))$ and we speak of \emph{$P$-rooted} forests. 
For an edge multiset $E_0$, let $\chi^{E_0}$ denote its multiplicity vector,
and let $\odd(E_0)$ be the set of vertices
of odd degree in $E_0$.
For $W\subseteq V$ of even size, a \emph{$W$-join} is an edge multiset
$J$ with $\odd(J)=W$, and $m(W)$ is the minimum $c$-cost of a $W$-join. 
Both $A$-rooted forests and $W$-joins can be computed at minimum cost in polynomial time.
% via a minimum spanning tree after identifying $A$ and a minimum-weight perfect matching on $W$, respectively.
By the theorem of Edmonds and
Johnson~\cite{EJ}, every $z\in\R^E_{\ge0}$ with $z(\delta(U))\ge1$ for all
$U\subseteq V$ with $|U\cap W|$ odd satisfies $m(W)\le c(z)$.

\paragraph{The relaxation.} 
We use the following undirected stroll
relaxation.
For distinct $s,t\in V$, the \emph{$s$--$t$ stroll polytope}
of~\cite{AMN} consists of all points
$(x,y)\in\R^E_{\ge0}\times\R^V_{\ge0}$ that satisfy
\[
 \begin{array}{r@{}l@{\qquad}l}
  y_s=y_t&{}=\frac12, &\\
  x(\delta(v))&{}=2y_v & \text{for all } v\in V,\\
  x(\delta(U))&{}\ge1 & \text{for all } U \text{ with } s\in U\subseteq V\setminus\{t\},\\
  x(\delta(U))&{}\ge2y_v & \text{for all } v \text{ and } U \text{ with } v\in U\subseteq V\setminus\{s,t\}.
 \end{array}
\]
The variables $x_e$ describe fractional edge usage, and $y_v$ indicates the \emph{coverage} of $v$.
For each pair $(s_i,t_i)$, we choose a point $(x^i,y^i)$ in its stroll
polytope. Our relaxation \LPrel{} minimizes $\sum_{i=1}^k c(x^i)$ subject to $\sum_{i=1}^k y^i_v=1$ for all $v \in N$, and $y^i_v=0$ for all  $v\in T\setminus\{s_i,t_i\}$.
That is, every non-terminal must be covered to a total extent of one, and each
stroll avoids all terminals other than its endpoints.
For Ordered TSP, this
is the relaxation of~\cite{AMN}.
For Multi-Path
TSP, \LPrel{} is equivalent to the flow relaxation of~\cite{BFMS}, restricted
so that the flow of each pair avoids the other terminals (Appendix~\ref{app:lp}). As
in~\cite{AMN}, \LPrel{} can be solved in polynomial time (Appendix~\ref{app:lp}). We fix an
optimal basic solution $(x^i,y^i)_{i\in[k]}$ of \LPrel{}, its value $\optlp$, and its aggregate
$X=\sum_ix^i\in\R^E_{\ge0}$. We have $\Delta\le\optlp\le\OPT$; Appendix~\ref{app:lp}
records the justification. The next lemma applies to each pair the tree decomposition
of~\cite[Lemma~5]{AMN}, which builds on Blauth and N\"agele~\cite[Lemma~4.2]{BN}; its
proof is in Appendix~\ref{app:lp}.

% Kept in a macro because Appendix~A restates it.
\newcommand{\LemDecomp}[1]{In polynomial time we can compute, for each $i\in[k]$, a
probability distribution#1 over trees $\Tr_i$ such that
(a)~$s_i,t_i\in\Tr_i$ for every $i$;
(b)~$\sum_i\Prob[v\in\Tr_i]=1$ for every $v\in N$;
(c)~$\sum_i\Prob[e\in\Tr_i]=X_e$ for every edge $e$.
Moreover, (d)~$X(\delta(U))\ge2$ for every nonempty $U\subseteq N$, and for Ordered TSP
even for every $U$ with $\emptyset\ne U\subsetneq V$.}%
\begin{lemma}\label{lem:decomp}
\LemDecomp{}
\end{lemma}

\paragraph{Trees, trunks, and branches.} From now on, $\Tr_1,\dots,\Tr_k$ are independent
random trees drawn from the distributions of Lemma~\ref{lem:decomp}. The \emph{trunk} $Q_i$
of $\Tr_i$ is its $s_i$--$t_i$ path, and the \emph{branches} $R_i$ are its other edges. Let
$Q=Q_1\uplus\dots\uplus Q_k$ and $R=R_1\uplus\dots\uplus R_k$ be the multisets of all trunk
and branch edges; thus $V(Q)=\bigcup_iV(Q_i)\supseteq T$ is the set of trunk vertices, and
$w(Q)=\sum_iw(Q_i)$ for $w\in\R^E_{\ge0}$. A $Q$-rooted forest is called a
\emph{connection forest}. Property~(c) gives $\E[w(Q)]+\E[w(R)]=w(X)$, and
the triangle inequality gives $w(Q)\ge c(Q)\ge\Delta$ whenever $w\ge c$. Solutions are
completed as follows (proof in Appendix~\ref{app:completion}).

% Kept in a macro because Appendix~B restates it.
\newcommand{\LemCompletion}{Let $K$ be a $Q$-rooted forest,
and let $J$ be an $\odd(K)$-join. In polynomial time we can compute $s_i$--$t_i$ paths, one
for every $i$, that together visit all vertices and cost at most $c(Q)+c(K)+c(J)$. For
Ordered TSP, we can instead compute a Hamiltonian cycle that visits $d_1,\dots,d_k$ in this
cyclic order, with the same bound on its cost.}%
\begin{lemma}\label{lem:completion}
\LemCompletion
\end{lemma}

\section{Connecting what the trunks miss}\label{sec:threshold}
Throughout this section, fix an arbitrary nonnegative edge-weight
function $w\in\R^E_{\ge0}$, which need not satisfy the triangle inequality.
For $t\ge0$, let $G_t=(V,\{e\in E:w_e\le t\})$.
Given a nonempty root set $A\subseteq V$, we call a vertex set $U \subseteq V$ 
\emph{rootless} if $U \cap A = \emptyset$.
Applying Kruskal's algorithm~\cite{Kruskal} to $G/A$ gives the following
integral formula for $f_w(A)$. The proof is in Appendix~\ref{app:tools}.

% Kept in a macro because Appendix~C restates it.
\newcommand{\LemKruskal}{For every nonempty root set $A\subseteq V$,
\[
  f_w(A)=\int_0^\infty
    \bigl|\{U: U \text{ is a rootless component of } G_t\}\bigr|\,dt .
\]
Moreover, $f_c(T)\le\optlp$.}%

\begin{lemma}\label{lem:kruskal}
\LemKruskal
\end{lemma}
\noindent 
Let $\Uc_w(t)$ be the family of terminal-free components of $G_t$.
In the following, we will only consider root sets $A$ which contain $T$. Thus, the rootless components of $G_t$ are all members of  $\Uc_w(t)$. Hence
\[
  f_w(A)=\int_0^\infty
    \bigl|\{U\in\Uc_w(t): U\cap A=\emptyset\}\bigr|\,dt .
\]
In particular, taking $A = T$ yields $f_w(T)=\int_0^\infty|\Uc_w(t)|\,dt$.

\paragraph{Warm-up: the Bridge Lemma.} The following lemma of B\"ohm et al.~\cite[Theorem~3]{BFMS} adapts
the Bridge Lemma of Byrka et al.~\cite{BGRS}.
We give a short proof using Lemma~\ref{lem:kruskal}, which also
prepares for its generalization in our Theorem~\ref{thm:threshold}.

\begin{lemma}[Bridge Lemma~\cite{BFMS}]\label{lem:bridge}
Let $S\subseteq N$ be a random set and $\gamma\ge0$ such that $\Prob[v\notin S]\le\gamma$ for
every $v\in N$. Then $\E[f_w(T\cup S)]\le\gamma\,f_w(T)$.
\end{lemma}

\begin{proof}
\Cref{lem:kruskal} for $A=T\cup S$ yields
$\E[f_w(T\cup S)]=\int_0^\infty\sum_{U\in\Uc_w(t)}\Prob[S\cap U=\emptyset]\,dt$. Each
$U\in\Uc_w(t)$ contains some vertex $v$, and $\Prob[S\cap U=\emptyset]\le\Prob[v\notin S]\le\gamma$.
We thus have $\E[f_w(T\cup S)]\le\int_0^\infty\gamma\,|\Uc_w(t)|\,dt=\gamma\,f_w(T)$.\qed
\end{proof}

\noindent Any $v\in N$ is missed by \emph{all} trees with probability $\prod_i(1-\Prob[v\in\Tr_i])\le\frac1e$. 
The analysis of \cite{BFMS} plugs this bound into the bridge lemma to derive $\E[f_c(Q \cup R)] \leq \frac 1 e f_c(T)$ as a bound on the connection cost. 
Recall that \cite{BFMS,AMN} keep the whole trees, whereas we only keep the trunks.
The following theorem shows that by doing so, we essentially trade each discarded unit of expected branch weight
for at most $1-\frac1e$ units of expected connection weight. 

\begin{figure}[t]
\centering
\begin{tikzpicture}[x=1.15cm,y=0.8cm,>={Latex[length=2.2mm]},
  term/.style={rectangle,draw=black,fill=white,minimum size=4.5mm,inner sep=0pt,font=\small},
  vtx/.style={circle,draw=black,fill=white,inner sep=1.4pt},
  inU/.style={circle,draw=black,fill=black!35,inner sep=1.4pt}]
  % the component U
  \fill[black!10,rounded corners=7pt] (4.15,0.95) rectangle (7.25,2.35);
  \draw[black!55,rounded corners=7pt] (4.15,0.95) rectangle (7.25,2.35);
  \node[font=\small] at (6.95,2.08) {$U$};
  % trunk
  \node[term] (s) at (0,0) {$s_i$};
  \node[term] (t) at (8,0) {$t_i$};
  \node[vtx] (a1) at (2,0) {};
  \node[vtx] (a2) at (3.8,0) {};
  \node[vtx] (a3) at (6,0) {};
  \draw[line width=1.6pt] (s) -- (a1) -- (a2) -- (a3) -- (t);
  \node[font=\small,anchor=north] at (5,-0.08) {trunk $Q_i$};
  % branches
  \node[vtx] (b1) at (1.4,1.35) {};
  \node[vtx] (b2) at (2.6,1.6) {};
  \draw[->,dashed,thick,black!70] (a1) -- (b1);
  \draw[->,dashed,thick,black!70] (b1) -- (b2);
  \node[inU] (u1) at (4.7,1.45) {};
  \node[inU] (u2) at (5.7,1.95) {};
  \node[inU] (u3) at (6.5,1.35) {};
  \draw[->,dashed,very thick,black] (a2) -- node[left,font=\small,pos=0.55] {$a$} (u1);
  \draw[->,dashed,thick,black!70] (u1) -- (u2);
  \draw[->,dashed,thick,black!70] (u2) -- (u3);
  \node[font=\small,anchor=west] at (0.15,2.05) {branches $R_i$};
\end{tikzpicture}
\caption{Charging in the proof of Theorem~\ref{thm:threshold}: trunk $Q_i$ (thick),
branches $R_i$ (dashed); the terminal-free component $U$ of $G_t$ is missed by
$Q_i$, and the branch arc $a$ enters $U$, so $t<w_a$.}
\label{fig:threshold}
\end{figure}
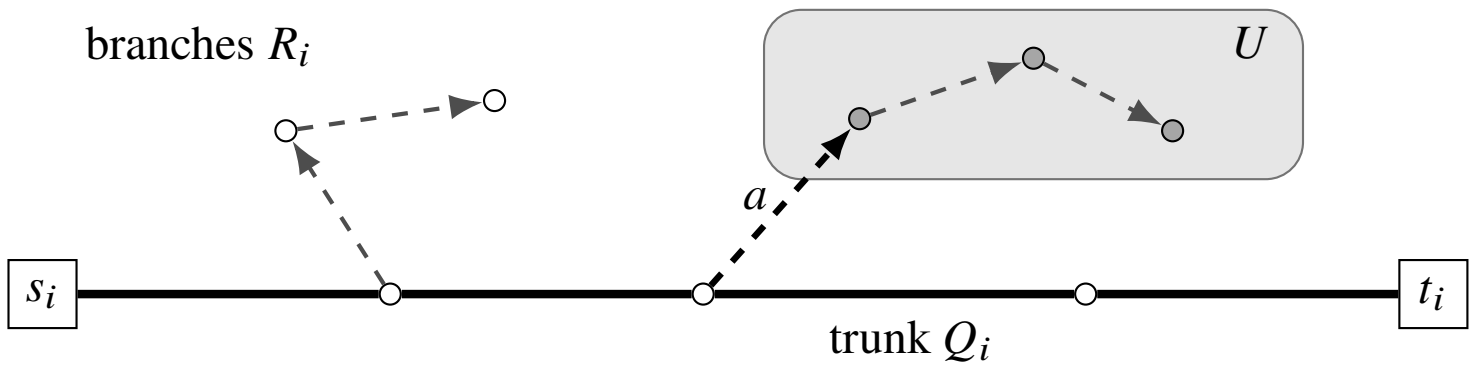

\begin{theorem}[Threshold connection bound]\label{thm:threshold}
The minimum $w$-cost $f_w(Q)$ of a connection forest satisfies
\[
 \E\bigl[f_w(Q)\bigr]\;\le\;\frac1e\,f_w(T)+\Bigl(1-\frac1e\Bigr)\,\E\bigl[w(R)\bigr].
\]
\end{theorem}

\begin{proof}
Orient every tree $\Tr_i$ away from $s_i$ (Figure~\ref{fig:threshold}). For $U\subseteq N$, let
\[
 \mu(U)=\sum_{i=1}^k\Prob\bigl[Q_i\cap U\ne\emptyset\bigr],\qquad
 \nu(U)=\sum_{i=1}^k\E\bigl[|R_i\cap\din(U)|\bigr],
\]
where $\din(U)$ is the set of arcs entering $U$.
Here $Q_i\cap U\ne\emptyset$ means that the trunk $Q_i$ visits a vertex
of $U$.
Thus, $\mu(U)$ is the expected number of trunks intersecting $U$, and $\nu(U)$ is the expected number of
branch arcs entering~$U$. 

We first show $\mu(U)+\nu(U)\ge1$ for every nonempty $U\subseteq N$. That is, in expectation $U$ has at least one incoming branch arc or intersecting trunk.
To this end, fix a vertex $v\in U$ and
an outcome of $\Tr_i$ such that $v \in \Tr_i$. The path in $\Tr_i$ from $s_i\notin U$ to $v \in U$ enters $U$ via some  arc $a$. If $Q_i$ is disjoint from $U$, then $a$ is a branch arc, i.e., $a \in R_i$. Hence
$\one[v\in\Tr_i]\le\one[Q_i\cap U\ne\emptyset]+|R_i\cap\din(U)|$. Taking expectations and
summing over $i$, Lemma~\ref{lem:decomp}(b) gives $1=\sum_i\Prob[v\in\Tr_i]\le\mu(U)+\nu(U)$.

Second, since the trees are independent and $1-p\le e^{-p}$,
\[
 \Prob[Q\cap U=\emptyset]=\prod_{i=1}^k\bigl(1-\Prob[Q_i\cap U\ne\emptyset]\bigr)
 \le e^{-\mu(U)}.
\]
The function $x\mapsto e^{-x}$ is convex on $[0,1]$ and at most $\frac1e$ on $[1,\infty)$,
so $e^{-x}\le\frac1e+(1-\frac1e)(1-x)^+$ for all $x\ge0$. Together with the first step,
$\Prob[Q\cap U=\emptyset]\le\frac1e+(1-\frac1e)\,\nu(U)$.

Now fix $t\ge0$. Each $U\in\Uc_w(t)$ is a nonempty subset of $N$, and an arc $a=(u,v)$
enters at most one of these sets, the one containing $v$, and none if $t\ge w_a$, as then
$u$ and $v$ lie in the same component. So summing the second step over $U\in\Uc_w(t)$
gives
\[
 \sum\nolimits_{U\in\Uc_w(t)}\Prob[Q\cap U=\emptyset]\;\le\;\frac1e\,|\Uc_w(t)|
 +\Bigl(1-\frac1e\Bigr)\sum\nolimits_i\E\bigl[|\{a\in R_i:w_a>t\}|\bigr].
\]
As $\Uc_w(t)$ does not depend on the trees, Lemma~\ref{lem:kruskal} for $A=V(Q)\supseteq T$
(in every outcome) and for $A=T$ shows that the first two terms integrate over $t$ to
$\E[f_w(Q)]$ and $\frac1ef_w(T)$; the last one integrates to
$(1-\frac1e)\sum_i\E[w(R_i)]=(1-\frac1e)\E[w(R)]$.\qed
\end{proof}
%The proof uses only that $Q_i$ is a subtree of $\Tr_i$ containing $s_i$ and $t_i$
%and that different trees are independent, not the triangle inequality for $w$. 
For both problems, we use the following construction and LP-relative bound as a candidate solution.

\begin{lemma}[Doubling]\label{lem:doubling}
Complete the trunks by Lemma~\ref{lem:completion} with a cheapest $Q$-rooted forest $K$
and set $J=K$. The cost
$\Adbl$ of the resulting solution satisfies
\begin{align*}
 \E[\Adbl]&\;\le\;2\Bigl(1-\frac1e\Bigr)\optlp+\frac2e\,f_c(T)-\Bigl(1-\frac2e\Bigr)\E[c(Q)]\\
 &\;\le\;2\,\optlp-\Bigl(1-\frac2e\Bigr)\E[c(Q)] .
\end{align*}
\end{lemma}

\begin{proof}
First, observe that $J$ is indeed a feasible $\odd(K)$-join. 
Thus, by Lemma~\ref{lem:completion}, the described construction yields a feasible solution of cost $\Adbl\le c(Q)+2f_c(Q)$. Theorem~\ref{thm:threshold}
with $w=c$ and $\E[c(R)]=\optlp-\E[c(Q)]$ gives
$\E[f_c(Q)]\le\frac1ef_c(T)+(1-\frac1e)(\optlp-\E[c(Q)])$, which proves the first
inequality. The second follows from $f_c(T)\le \optlp$ (Lemma~\ref{lem:kruskal}).\qed
\end{proof}
We use the derived bound to prove Corollary~\ref{cor:gap-mp}.
\begin{proof}[of Corollary~\ref{cor:gap-mp}]
Some outcome of the trees gives a feasible solution of cost at most $\E[\Adbl]$, and
$c(Q)\ge\Delta$; hence $\OPT\le2\,\optlp-(1-\frac2e)\Delta$ by Lemma~\ref{lem:doubling}.
The other bound follows from Theorem~\ref{thm:mp}.\qed
\end{proof}

\section{Ordered TSP}\label{sec:ot}

For Ordered TSP, half of the LP solution pays for any parity correction, and
Theorem~\ref{thm:threshold} with $w=c$ suffices. The randomized algorithm solves
\LPrel{}, computes the tree distributions of Lemma~\ref{lem:decomp}, draws the trees
independently, and computes two tours: the \emph{doubling} tour of
Lemma~\ref{lem:doubling} of cost $\Adbl$, and the \emph{joining} tour, which completes the trunks by
Lemma~\ref{lem:completion} with a cheapest $Q$-rooted forest $K$ and a cheapest
$\odd(K)$-join $J$; let $\Ajoin$ be its cost. It returns the cheaper tour.
For fixed trunks and $K$, the minimum join costs no more than $K$ itself, so both
bounds below apply to the joining completion. Their separate derandomizations may
choose different trunks.

\begin{lemma}\label{lem:ot}
The randomized algorithm returns a tour of expected cost at most $\rhoOT\,\optlp$.
\end{lemma}

\begin{proof}
By property~(d) of Lemma~\ref{lem:decomp}, $\frac12X(\delta(U))\ge1$ for every $U$ with
$\emptyset\ne U\subsetneq V$, so $m(W)\le\frac12c(X)=\frac12\,\optlp$ for every even set $W$ by the
theorem of Edmonds and Johnson. Hence $\Ajoin\le c(Q)+f_c(Q)+\frac12\,\optlp$, and
Theorem~\ref{thm:threshold} with $w=c$ and $\E[c(R)]=\optlp-\E[c(Q)]$ gives
\[
 \E[\Ajoin]\;\le\;\Bigl(\frac32-\frac1e\Bigr)\optlp+\frac1e\,f_c(T)+\frac1e\,\E[c(Q)] .
\]
The expected cost of the cheaper tour is at most the combination of the bound of
Lemma~\ref{lem:doubling} and this bound with the weights $\frac1{e-1}$ and
$\frac{e-2}{e-1}$, in which $\E[c(Q)]$ cancels:
\[
 \E\bigl[\min\{\Adbl,\Ajoin\}\bigr]\;\le\;\frac{3e-4}{2(e-1)}\,\optlp+\frac1{e-1}\,f_c(T)
 \;\le\;\Bigl(1+\frac{e}{2(e-1)}\Bigr)\optlp,
\]
by $f_c(T)\le \optlp$; Appendix~\ref{app:constants} verifies the decimal bound.\qed
\end{proof}

\noindent Corollary~\ref{cor:gap} follows, because some outcome of the trees gives a tour
of at most this cost. Choosing the trunks by the method of conditional expectations
instead of at random proves Theorem~\ref{thm:ot} (Appendix~\ref{app:tools}).

\section{Labels, parity, and the reweighting theorem}\label{sec:growth}

\paragraph{The growth process.} Fix an instance of Multi-Path TSP and $\lambda\in[0,1)$.
We run the growth process of~\cite{ZLX} with the terminals as roots. The process  maintains a forest $F$ on $V$. At time $t$, let $\Ac_t$ be the family of rootless components
of $F$ at time $t$, and let $U_t$ be their union. 
A vertex is
\emph{rooted} once its component contains a terminal. 
Furthermore, the process maintains dual variables
$y_U\ge0$ for $U\subseteq N$.
At time $t=0$, the forest is empty, and all dual variables are zero. 
Thus $\Ac_0$ consists of the singletons $\{v\}$ for $v\in N$,
and $U_0=N$.

At time $t$ it raises $y_U$ at rate
$\frac12$ for every $U\in\Ac_t$ and $y_{U_t}$ at rate $\frac{1-\lambda}{2(1+\lambda)}$ (the
rates add if $U_t\in\Ac_t$). The \emph{load} of an edge $e$ at time $t$ is
$\zeta_e(t)=\sum_{U:e\in\delta(U)}y_U$, and $\zeta_e$ denotes its final value. 
When an edge becomes tight (when $\zeta_e(t) = c(e)$), add it to $F$ if its endpoints are in
distinct components of $F/T$, processing ties one at a time.
Stop when all vertices are rooted.
The \emph{label} $h(v)$ is the time at
which $v$ becomes rooted, so $h(v)=0$ for $v\in T$. We call $F$ the \emph{source forest},
and we define the \emph{surcharge} and the \emph{reweighted cost}
\[
 \theta_{uv}=\bigl(c_{uv}-\min\{h(u),h(v)\}\bigr)^+,\qquad w=c+\lambda\theta .
\]
For rational $\lambda$, the process runs in polynomial time (Appendix~\ref{app:growth}). As $\zeta\le c$ (Appendix~\ref{app:growth}), $y$ is a
feasible solution of the dual of the cut relaxation
$\min\{c(x):x(\delta(U))\ge2\text{ for all }\emptyset\ne U\subseteq N,\ x\ge0\}$.

The load of an edge joining two distinct rootless components grows at
rate $1$.
While exactly one endpoint is rooted, the load grows at rate  
$\frac12+\frac{1-\lambda}{2(1+\lambda)}=\frac1{1+\lambda}$. 
Using these rates, a simple calculation shows that every edge $e$ added to $F$ at time $t$ has $w_e=t$. Together with
a cycle argument, this shows that $F$
is a minimum $w$-cost $T$-rooted forest.
For details, see Appendix~\ref{app:growth}.

We define the \emph{fractional sum of peaks} by
\[
  H=\sum_{i=1}^k \int_0^\infty \max_{v\in U_t} y^i_v\,dt,
\]
where the maximum over the empty set is zero. For an integral assignment,
the $i$th maximum is one as long as the $i$th path contains an unrooted vertex.
The $i$th integral is therefore the largest label $h(v)$ on the $i$th path.

% Kept in a macro because Appendix~D restates it.
\newcommand{\LemSource}{Let $0\le\lambda<1$.
(a)~The source forest $F$ is a minimum $w$-cost $T$-rooted forest, and
$w(F)=f_w(T)=\int_0^\infty|\Ac_t|\,dt$.
(b)~$f_w(T)+\frac{1-\lambda}{1+\lambda}H\le\optlp$.}%
\begin{lemma}\label{lem:source}
\LemSource
\end{lemma}

\noindent Part~(a) is~\cite[Lemma~2]{ZLX}. For part~(b), we apply the load
calculation of~\cite[Lemma~4]{ZLX} to $X$. Each rootless component
has cut value at least two. For their union $U_t$, the stroll cut
constraints give a contribution from each pair, which yields the
corresponding term of $H$ after integration
(Appendix~\ref{app:growth}).

% Kept in a macro because Appendix~D restates it.
\newcommand{\LemParity}{For every parameter $\lambda\in[0,1)$ and every
multiset $K$ of edges,
\[
 m(\odd(K))\;\le\;\tfrac12(\optlp+\Delta)+H+\theta(K).
\]}%
\begin{lemma}\label{lem:parity}
\LemParity
\end{lemma}

\noindent This is the LP-relative counterpart of~\cite[Lemma~3]{ZLX}, whose
bound uses the sum of the largest labels on the tours of an optimal
solution. Half of $X$ and half of the direct pair edges account for
the first term. To cover the remaining deficient cuts, we use selected
fractions of the edges of $K$. The packing inequalities bound their
label cost by $H$ after integration over $U_t$, and $\theta(K)$ covers
the remaining cost. Appendix~\ref{app:growth} gives the proof.

Section~\ref{sec:mp} applies Theorem~\ref{thm:threshold} to $w$ and therefore needs a
bound on $w(X)$.

\begin{theorem}[Reweighting paid by dual slack]\label{thm:reweight}
Let $0\le\lambda<1$, and let $w=c+\lambda\theta$ be given by the growth process with
parameter~$\lambda$. Then
\[
 f_w(T)\le \optlp\qquad\text{and}\qquad w(X)\;\le\;\optlp+\frac{2\lambda}{1-\lambda}\,\bigl(\optlp-f_w(T)\bigr).
\]
\end{theorem}

\begin{proof}
The surcharge of an edge is large only if its cost exceeds the label of its earlier
endpoint by much, in which case the edge carries dual slack or lies across the slowly
growing set $U_t$ for a long time; we show that both are paid for by $\optlp-f_w(T)$. For an
edge $uv$ with $h(u)\le h(v)$, the load grows at rate at most $1$ before time $h(u)$, at
rate at most $\frac1{1+\lambda}$ while exactly one endpoint is rooted, and not after time
$h(v)$. Hence $\zeta_{uv}\le h(u)+\frac{h(v)-h(u)}{1+\lambda}$. If $c_{uv}\ge h(u)$, this gives
$\theta_{uv}=c_{uv}-h(u)\le(c_{uv}-\zeta_{uv})+\frac{h(v)-h(u)}{1+\lambda}$; otherwise
$\theta_{uv}=0\le c_{uv}-\zeta_{uv}$. Thus every edge satisfies
\begin{equation}\label{eq:envelope}
 \theta_{uv}\;\le\;c_{uv}-\zeta_{uv}+\frac{|h(u)-h(v)|}{1+\lambda} .
\end{equation}
The final loads are the integrals of the rates at which they grow, an edge $uv$ lies in
$\delta(U_t)$ exactly when $\min\{h(u),h(v)\}\le t<\max\{h(u),h(v)\}$, and every
$U\in\Ac_t$ is a nonempty subset of $N$, so that $X(\delta(U))\ge2$ by
Lemma~\ref{lem:decomp}(d). With Lemma~\ref{lem:source}(a) and $\zeta\le c$, this gives
\begin{align}
 \optlp\;\ge\;\sum_eX_e\zeta_e
 &\;=\;\int_0^\infty\Bigl(\frac12\sum_{U\in\Ac_t}X(\delta(U))+\frac{1-\lambda}{2(1+\lambda)}\,X(\delta(U_t))\Bigr)dt\notag\\
 &\;\ge\;f_w(T)+\frac{1-\lambda}{2(1+\lambda)}\sum_{uv}X_{uv}|h(u)-h(v)| ,\label{eq:loadbudget}
\end{align}
and in particular $f_w(T)\le \optlp$. Multiplying~\eqref{eq:envelope} by $X_{uv}$, summing,
and using~\eqref{eq:loadbudget} twice, we get, as
$\frac1{1+\lambda}-\frac{1-\lambda}{2(1+\lambda)}=\frac12$ and $1+\frac{1+\lambda}{1-\lambda}=\frac2{1-\lambda}$,
\[
 \theta(X)\;\le\;\optlp-f_w(T)+\frac12\sum_{uv}X_{uv}|h(u)-h(v)|
 \;\le\;\frac2{1-\lambda}\bigl(\optlp-f_w(T)\bigr).
\]
As $w(X)=\optlp+\lambda\theta(X)$, the theorem follows.\qed
\end{proof}

\section{Multi-Path TSP}\label{sec:mp}

The randomized algorithm uses $\lambda=\frac2{15}$. It solves \LPrel{}, computes the tree
distributions of Lemma~\ref{lem:decomp}, draws the trees independently, and runs the growth process with parameter
$\lambda$ to obtain $w=c+\lambda\theta$. It computes
two solutions from their trunks and returns the cheaper one: the \emph{reweighted
candidate}, which completes the trunks by Lemma~\ref{lem:completion} with a cheapest
$Q$-rooted forest $K$ with respect to $w$ and a cheapest $\odd(K)$-join $J$ with
respect to $c$, at cost $\Alam$; and the \emph{doubling candidate} of
Lemma~\ref{lem:doubling}, at cost $\Adbl$. Here the two connection forests are
optimized in different weights, $w$ and $c$, respectively.

For every forest $K$, the
trivial bound $m(\odd(K))\le c(K)$ with weight $\frac{1-\lambda}{1+\lambda}$ and
Lemma~\ref{lem:parity} with weight $\frac{2\lambda}{1+\lambda}$ give
\begin{equation}\label{eq:combination}
 c(K)+m(\odd(K))\;\le\;\frac2{1+\lambda}\,w(K)+\frac{2\lambda}{1+\lambda}\Bigl(\frac{\optlp+\Delta}2+H\Bigr),
\end{equation}
since $c(K)+\frac{1-\lambda}{1+\lambda}c(K)+\frac{2\lambda}{1+\lambda}\theta(K)=\frac2{1+\lambda}w(K)$. As the
last term does not depend on $K$, the reweighted candidate optimizes the forest in $w$.
The next lemma bounds its expected cost for a general parameter.

\begin{lemma}\label{lem:master}
Let $0<\lambda\le1-\frac2e$, and let
\[
 \kappa_\lambda=\frac{2\lambda-1+\frac2e}{1+\lambda},\qquad
 \pi_\lambda=\frac{2}{1-\lambda^2}\Bigl(\frac{1+\lambda}e-3\lambda-\lambda^2\Bigr).
\]
Then the cost $\Alam$ of the reweighted candidate, computed with parameter $\lambda$,
satisfies
\[
 \E[\Alam]\;\le\;
 \left(\frac{2(1-\frac1e)}{1-\lambda}+\frac{\lambda(3+\lambda)}{1-\lambda^2}\right)\optlp
 +\kappa_\lambda\Delta+\pi_\lambda f_w(T).
\]
\end{lemma}

\begin{proof}
By Lemma~\ref{lem:completion} and~\eqref{eq:combination} with $w(K)=f_w(Q)$, the paths
cost at most $c(Q)+\frac2{1+\lambda}f_w(Q)+\frac{2\lambda}{1+\lambda}(\frac{\optlp+\Delta}2+H)$.
Theorem~\ref{thm:threshold} bounds $\E[f_w(Q)]$, and with
$\alpha=\frac{2(1-1/e)}{1+\lambda}$, which is at least $1$ because $\lambda\le1-\frac2e$,
\[
 \E[\Alam]\;\le\;\E[c(Q)]+\alpha\,\E[w(R)]+\frac2{e(1+\lambda)}\,f_w(T)
 +\frac{2\lambda}{1+\lambda}\Bigl(\frac{\optlp+\Delta}2+H\Bigr).
\]
Since $w\ge c$, we have $c(Q)\le w(Q)$ and $w(Q)\ge\Delta$, so $\E[w(Q)]+\E[w(R)]=w(X)$ gives
\[
 \E[c(Q)]+\alpha\,\E[w(R)]\;\le\;\alpha\,w(X)-(\alpha-1)\,\E[w(Q)]\;\le\;\alpha\,w(X)-(\alpha-1)\Delta .
\]
Theorem~\ref{thm:reweight} gives $w(X)\le \optlp+\frac{2\lambda}{1-\lambda}(\optlp-f_w(T))$, and
Lemma~\ref{lem:source}(b) gives $\frac{2\lambda}{1+\lambda}H\le\frac{2\lambda}{1-\lambda}(\optlp-f_w(T))$.
Hence $\E[\Alam]$ is at most
\begin{align*}
 &\frac{\alpha(1+\lambda)}{1-\lambda}\,\optlp+\Bigl(\frac\lambda{1+\lambda}+\frac{2\lambda}{1-\lambda}\Bigr)\optlp\\
 &\quad+\Bigl(1-\alpha+\frac\lambda{1+\lambda}\Bigr)\Delta
 +\Bigl(\frac2{e(1+\lambda)}-\frac{2\lambda(1+\alpha)}{1-\lambda}\Bigr)f_w(T),
\end{align*}
and collecting the coefficients gives the lemma.\qed
\end{proof}

\begin{lemma}\label{lem:mp}
The randomized algorithm returns a solution of expected cost at most $\rhoMP\,\optlp$.
\end{lemma}

\begin{proof}
For $\lambda=\frac2{15}$, we have $\kappa_\lambda=\frac1{17}(\frac{30}e-11)>0$ and
$\pi_\lambda=-\frac1{221}(188-\frac{510}e)<0$, and Lemma~\ref{lem:master} with
$f_w(T)\ge f_c(T)$ (as $w\ge c$) gives (Appendix~\ref{app:constants})
\[
 \E[\Alam]\;\le\;\frac{32}{17}\,\optlp+\frac1{17}\Bigl(\frac{30}e-11\Bigr)\Delta
 +\frac1{221}\Bigl(188-\frac{510}e\Bigr)\bigl(\optlp-f_c(T)\bigr).
\]
Lemma~\ref{lem:doubling} and $\E[c(Q)]\ge\Delta$ give
\[
 \E[\Adbl]\;\le\;2\,\optlp-\Bigl(1-\frac2e\Bigr)\Delta-\frac2e\bigl(\optlp-f_c(T)\bigr).
\]
The expected cost of the cheaper candidate is at most the convex combination of these
bounds with the weight $\frac{17(e-2)}{6e-4}$ on the first and $\frac{30-11e}{6e-4}$ on the
second; both are positive. In this combination, the coefficient of $\Delta$ vanishes, the coefficient of
$\optlp$ is $\frac{5e-2}{3e-2}$, and the coefficient of $\optlp-f_c(T)$ is
$-\frac{2(150e-47e^2-60)}{13e(3e-2)}<0$ (Appendix~\ref{app:constants}). Since
$\optlp\ge f_c(T)$ (Lemma~\ref{lem:kruskal}), the lemma
follows.\qed
\end{proof}

\noindent Choosing the trunks by the method of conditional expectations instead of at random
proves Theorem~\ref{thm:mp} (Appendix~\ref{app:tools}).

%\section{Open questions}\label{sec:open}
%What are the exact integrality gaps of \LPrel{} and of the Ordered-TSP relaxation
%of~\cite{AMN}, and are our analyses tight? The commodity coverage information is
%essential to our LP-relative parity bound: the weaker multiple-depot cut relaxation
%of Section~\ref{sec:growth} has integrality gap at least~$2$~\cite[Remark~1]{ZLX}.
%Does Theorem~\ref{thm:threshold} improve the ratios for the prize-collecting
%versions~\cite{AMR}?
\label{end-of-main}
\begin{credits}
\subsubsection{\ackname} 
Generative AI tools, including ChatGPT and Claude, were used extensively throughout this work.  They
contributed to the formulation of theorems, algorithms, proofs, and other
claims.  They were also used for literature search and citation preparation,
figure production, and the drafting and iterative
revision of the manuscript.  The authors guided and reviewed this process.
The authors have checked all content and assume full responsibility for it. % the content.
\end{credits}
\bibliographystyle{splncs04}
\bibliography{current}

@inproceedings{AMN,
  author    = {Susanne Armbruster and Matthias Mnich and Martin N{\"a}gele},
  title     = {A $(3/2 + 1/e)$-Approximation Algorithm for Ordered {TSP}},
  booktitle = {Approximation, Randomization, and Combinatorial Optimization. Algorithms and Techniques (APPROX/RANDOM 2024)},
  editor    = {Amit Kumar and Noga Ron-Zewi},
  series    = {Leibniz International Proceedings in Informatics (LIPIcs)},
  volume    = {317},
  pages     = {1:1--1:18},
  publisher = {Schloss Dagstuhl -- Leibniz-Zentrum f{\"u}r Informatik},
  address   = {Dagstuhl, Germany},
  year      = {2024},
  doi       = {10.4230/LIPIcs.APPROX/RANDOM.2024.1}
}

@inproceedings{BFMS,
  author    = {Martin B{\"o}hm and Zachary Friggstad and Tobias M{\"o}mke and Joachim Spoerhase},
  title     = {Approximating Traveling Salesman Problems Using a Bridge Lemma},
  booktitle = {Proceedings of the 2025 Annual ACM-SIAM Symposium on Discrete Algorithms (SODA)},
  pages     = {1166--1177},
  publisher = {SIAM},
  address   = {Philadelphia, PA},
  year      = {2025},
  doi       = {10.1137/1.9781611978322.34}
}

@misc{ZLX,
  author        = {Jingyang Zhao and Yuxi Liu and Mingyu Xiao},
  title         = {A $(1+1/\sqrt{2})$-Approximation for the Multiple-Depot Traveling Salesman Problem},
  howpublished  = {arXiv preprint arXiv:2609.19537},
  eprint        = {2609.19537},
  archivePrefix = {arXiv},
  primaryClass  = {cs.DS},
  year          = {2026}
}

@inproceedings{ADMPV,
  author    = {Morteza Alimi and Niklas Dahlmeier and Tobias M{\"o}mke and Philipp Pabst and Laura {Vargas Koch}},
  title     = {Approximating Graphic Multi-Path {TSP} and Graphic Ordered {TSP}},
  booktitle = {2026 SIAM Symposium on Simplicity in Algorithms (SOSA)},
  pages     = {524--537},
  publisher = {SIAM},
  address   = {Philadelphia, PA},
  year      = {2026},
  doi       = {10.1137/1.9781611978964.43}
}

@misc{DLPV,
  author        = {Niklas Dahlmeier and Raoul Luqu{\'e} and Philipp Pabst and Laura {Vargas Koch}},
  title         = {Approximation Algorithms for Graphic Multi-Depot {TSP} and Graphic Multi-Path {TSP}},
  howpublished  = {arXiv preprint arXiv:2610.16893},
  eprint        = {2610.16893},
  archivePrefix = {arXiv},
  primaryClass  = {cs.DS},
  year          = {2026}
}

@inproceedings{AMR,
  author    = {Morteza Alimi and Tobias M{\"o}mke and Michael Ruderer},
  title     = {Approximating Prize-Collecting Variants of {TSP}},
  booktitle = {50th International Symposium on Mathematical Foundations of Computer Science (MFCS 2025)},
  editor    = {Pawe{\l} Gawrychowski and Filip Mazowiecki and Micha{\l} Skrzypczak},
  series    = {Leibniz International Proceedings in Informatics (LIPIcs)},
  volume    = {345},
  pages     = {7:1--7:17},
  publisher = {Schloss Dagstuhl -- Leibniz-Zentrum f{\"u}r Informatik},
  address   = {Dagstuhl, Germany},
  year      = {2025},
  doi       = {10.4230/LIPIcs.MFCS.2025.7}
}

@article{BFJ,
  author  = {J{\o}rgen Bang-Jensen and Andr{\'a}s Frank and Bill Jackson},
  title   = {Preserving and Increasing Local Edge-Connectivity in Mixed Graphs},
  journal = {SIAM Journal on Discrete Mathematics},
  volume  = {8},
  number  = {2},
  pages   = {155--178},
  year    = {1995},
  doi     = {10.1137/S0036142993226983}
}

@article{BGRS,
  author  = {Jaros{\l}aw Byrka and Fabrizio Grandoni and Thomas Rothvo{\ss} and Laura Sanit{\`a}},
  title   = {Steiner Tree Approximation via Iterative Randomized Rounding},
  journal = {Journal of the ACM},
  volume  = {60},
  number  = {1},
  pages   = {6:1--6:33},
  year    = {2013},
  doi     = {10.1145/2432622.2432628}
}

@article{EJ,
  author  = {Jack Edmonds and Ellis L. Johnson},
  title   = {Matching, {E}uler Tours and the {C}hinese Postman},
  journal = {Mathematical Programming},
  volume  = {5},
  number  = {1},
  pages   = {88--124},
  year    = {1973},
  doi     = {10.1007/BF01580113}
}

@techreport{Christofides,
  author      = {Nicos Christofides},
  title       = {Worst-Case Analysis of a New Heuristic for the Travelling Salesman Problem},
  institution = {Graduate School of Industrial Administration, Carnegie-Mellon University},
  address     = {Pittsburgh, PA},
  type        = {Technical Report},
  number      = {388},
  month       = feb,
  year        = {1976},
  note        = {Reprinted in \emph{Operations Research Forum} 3(1), Article 20, 2022},
  doi         = {10.1007/s43069-021-00101-z}
}

@article{Serdyukov,
  author  = {A. I. Serdyukov},
  title   = {O nekotorykh ekstremal'nykh obkhodakh v grafakh},
  journal = {Upravlyaemye Sistemy},
  volume  = {17},
  pages   = {76--79},
  year    = {1978},
  note    = {In Russian. English title: On some extremal walks in graphs. English translation in R.~van Bevern and V.~A. Slugina, \emph{Historia Mathematica} 53:118--127, 2020}
}

@book{TV,
  author    = {Vera Traub and Jens Vygen},
  title     = {Approximation Algorithms for Traveling Salesman Problems},
  publisher = {Cambridge University Press},
  address   = {Cambridge, UK},
  year      = {2025},
  isbn      = {978-1-009-44541-2},
  doi       = {10.1017/9781009445436}
}

@inproceedings{Zenklusen,
  author    = {Rico Zenklusen},
  title     = {A 1.5-Approximation for Path {TSP}},
  booktitle = {Proceedings of the Thirtieth Annual ACM-SIAM Symposium on Discrete Algorithms (SODA 2019)},
  pages     = {1539--1549},
  publisher = {SIAM},
  address   = {Philadelphia, PA},
  year      = {2019},
  doi       = {10.1137/1.9781611975482.93}
}

@article{GW,
  author  = {Michel X. Goemans and David P. Williamson},
  title   = {A General Approximation Technique for Constrained Forest Problems},
  journal = {SIAM Journal on Computing},
  volume  = {24},
  number  = {2},
  pages   = {296--317},
  year    = {1995},
  doi     = {10.1137/S0097539793242618}
}

@article{Kruskal,
  author  = {Joseph B. Kruskal},
  title   = {On the Shortest Spanning Subtree of a Graph and the Traveling Salesman Problem},
  journal = {Proceedings of the American Mathematical Society},
  volume  = {7},
  number  = {1},
  pages   = {48--50},
  year    = {1956},
  doi     = {10.1090/S0002-9939-1956-0078686-7}
}

@article{Wolsey,
  author  = {Laurence A. Wolsey},
  title   = {Heuristic Analysis, Linear Programming and Branch and Bound},
  journal = {Mathematical Programming Study},
  volume  = {13},
  pages   = {121--134},
  year    = {1980},
  doi     = {10.1007/BFb0120913}
}

@article{SW,
  author  = {David B. Shmoys and David P. Williamson},
  title   = {Analyzing the {Held-Karp} {TSP} Bound: A Monotonicity Property with Application},
  journal = {Information Processing Letters},
  volume  = {35},
  number  = {6},
  pages   = {281--285},
  year    = {1990},
  doi     = {10.1016/0020-0190(90)90028-V}
}

@inproceedings{BHKK,
  author    = {Hans-Joachim B{\"o}ckenhauer and Juraj Hromkovi{\v{c}} and Joachim Kneis and Joachim Kupke},
  title     = {On the Approximation Hardness of Some Generalizations of {TSP}},
  booktitle = {Algorithm Theory -- SWAT 2006: 10th Scandinavian Workshop on Algorithm Theory},
  editor    = {Lars Arge and Rusins Freivalds},
  series    = {Lecture Notes in Computer Science},
  volume    = {4059},
  pages     = {184--195},
  publisher = {Springer},
  address   = {Berlin, Heidelberg},
  year      = {2006},
  doi       = {10.1007/11785293_19}
}

@article{BMS,
  author  = {Hans-Joachim B{\"o}ckenhauer and Tobias M{\"o}mke and Monika Steinov{\'a}},
  title   = {Improved Approximations for {TSP} with Simple Precedence Constraints},
  journal = {Journal of Discrete Algorithms},
  volume  = {21},
  pages   = {32--40},
  year    = {2013},
  doi     = {10.1016/j.jda.2013.04.002}
}

@inproceedings{BN,
  author    = {Jannis Blauth and Martin N{\"a}gele},
  title     = {An Improved Approximation Guarantee for Prize-Collecting {TSP}},
  booktitle = {Proceedings of the 55th Annual ACM Symposium on Theory of Computing (STOC 2023)},
  pages     = {1848--1861},
  publisher = {ACM},
  address   = {New York, NY, USA},
  year      = {2023},
  doi       = {10.1145/3564246.3585159}
}

@inproceedings{DKM,
  author    = {Max Deppert and Matthias Kaul and Matthias Mnich},
  title     = {A $(3/2+\varepsilon)$-Approximation for Multiple {TSP} with a Variable Number of Depots},
  booktitle = {31st Annual European Symposium on Algorithms (ESA 2023)},
  editor    = {Inge Li G{\o}rtz and Martin Farach-Colton and Simon J. Puglisi and Grzegorz Herman},
  series    = {Leibniz International Proceedings in Informatics (LIPIcs)},
  volume    = {274},
  pages     = {39:1--39:15},
  publisher = {Schloss Dagstuhl -- Leibniz-Zentrum f{\"u}r Informatik},
  address   = {Dagstuhl, Germany},
  year      = {2023},
  doi       = {10.4230/LIPIcs.ESA.2023.39}
}

@article{XR,
  author  = {Zhou Xu and Brian Rodrigues},
  title   = {A $3/2$-Approximation Algorithm for the Multiple {TSP} with a Fixed Number of Depots},
  journal = {INFORMS Journal on Computing},
  volume  = {27},
  number  = {4},
  pages   = {636--645},
  year    = {2015},
  doi     = {10.1287/ijoc.2015.0650}
}
\appendix
\renewcommand{\theHsection}{appendix.\Alph{section}}

\section{\texorpdfstring{The relaxation: proof of Lemma~\ref{lem:decomp}, and the flow relaxation}{The relaxation}}\label{app:lp}

\begin{restate}[restated in full]{Lemma~\ref{lem:decomp}}
\LemDecomp{ with polynomial support and rational probabilities}
\end{restate}

\begin{proof}
For each $i$, the point $(x^i,y^i)$ lies in the $s_i$--$t_i$ stroll polytope.
Lemma~5 of~\cite{AMN}, which derives from~\cite[Lemma~4.2]{BN}, computes for it in
polynomial time a family of trees $B$ that contain $s_i$ and $t_i$, with nonnegative
rational weights $\omega_B$ summing to one, such that $\sum_B\omega_B\chi^{E(B)}=x^i$ and
$\sum_{B\ni v}\omega_B=y^i_v$ for every $v\notin\{s_i,t_i\}$. Let $\Tr_i$ take the value $B$
with probability $\omega_B$; this gives~(a). For $v\in N$,
$\sum_i\Prob[v\in\Tr_i]=\sum_iy^i_v=1$, which is~(b), and
$\sum_i\Prob[e\in\Tr_i]=\sum_ix^i_e=X_e$, which is~(c).

For~(d), let $\emptyset\ne U\subseteq N$ and $v\in U$. For every $i$, we have
$U\subseteq V\setminus\{s_i,t_i\}$, so $x^i(\delta(U))\ge2y^i_v$, and summing over $i$ gives
$X(\delta(U))\ge2$. For Ordered TSP, $X$ even lies in the subtour elimination
polytope~\cite[Observation~3]{AMN}; we repeat the argument. Let
$\emptyset\ne U\subsetneq V$. If $U\cap T=\emptyset$ or $T\subseteq U$, the previous
argument applies to $U$ or to $V\setminus U$, which has the same cut. Otherwise both $U$
and $V\setminus U$ contain ordered vertices, so for at least two indices $i$ exactly one of
$d_i,d_{i+1}$ lies in $U$; for each of them, $x^i(\delta(U))\ge1$ by the cut constraint of
the stroll polytope applied to $U$ or to $V\setminus U$. Hence $X(\delta(U))\ge2$.\qed
\end{proof}

\paragraph{Cost bounds.} We verify $\Delta\le\optlp\le\OPT$, as stated in
Section~\ref{sec:prelim}. By Lemma~\ref{lem:decomp}(a), every $\Tr_i$ contains an $s_i$--$t_i$ path of cost at least
$c(s_i,t_i)$, and part~(c) gives $\sum_i\E[c(\Tr_i)]=c(X)=\optlp$; hence $\Delta\le\optlp$. Finally, for Multi-Path TSP, let
$x^i$ be the incidence vector of the optimal path $O_i$ fixed in Section~\ref{sec:prelim},
$y^i_v=1$ for the inner vertices $v$ of $O_i$, $y^i_{s_i}=y^i_{t_i}=\frac12$, and $y^i_v=0$
otherwise. Since a path from $s_i$ to $t_i$ crosses every cut that separates $s_i$ from
$t_i$, and enters and leaves every set $U\subseteq V\setminus\{s_i,t_i\}$ that contains one of
its vertices, these are points of the stroll polytopes; as every vertex of $N$ lies on
exactly one $O_i$ and no $O_i$ contains another terminal, they are feasible for \LPrel{},
so $\optlp\le\OPT$. For Ordered TSP, the same applies to the $k$ subpaths of an optimal tour
between consecutive ordered vertices.

\paragraph{Solving \LPrel{}.} For fixed $i$ and $v\in V\setminus\{s_i,t_i\}$, the
constraints $x^i(\delta(U))\ge2y^i_v$ for $v\in U\subseteq V\setminus\{s_i,t_i\}$ say that
every cut separating $v$ from $\{s_i,t_i\}$ has capacity at least $2y^i_v$ under the
capacities $x^i$. By the max-flow min-cut theorem, they are equivalent to the existence of
a flow of value $2y^i_v$ from $v$ to $\{s_i,t_i\}$ within these capacities; similarly, the
cut constraints for the sets $U$ with $s_i\in U\subseteq V\setminus\{t_i\}$ are equivalent to
an $s_i$--$t_i$ flow of value~$1$. Introducing these flows as variables gives an
equivalent linear program of polynomial size, which can be solved exactly in polynomial
time; an optimal basic solution has polynomial encoding length. (Armbruster et
al.~\cite{AMN} instead separate the cut constraints by minimum-cut computations.)

\paragraph{The flow relaxation.} For Multi-Path TSP, let $D_i$ be the digraph on
$V(D_i)=N\cup\{s_i,t_i\}$ that contains all arcs between these vertices except those that
enter $s_i$ or leave $t_i$, and let $\dout(U)$ and $\din(U)$ denote the sets of arcs of
$D_i$ that leave and enter $U\subseteq V(D_i)$. The relaxation (LP-kTSPP) of B\"ohm et
al.~\cite{BFMS}, restricted to flows $x_i$ of each pair $i$ in $D_i$, minimizes
$\sum_i\sum_{a\in D_i}c(a)\,x_{i,a}$ subject to $x,z\ge0$, $\sum_iz_{iv}=1$ for $v\in N$,
$x_i(\dout(s_i))=x_i(\din(t_i))=1$ and $x_i(\din(v))=x_i(\dout(v))=z_{iv}$ for $v\in N$, and
$x_i(\din(U))\ge z_{iv}$ for $v\in N$ and $v\in U\subseteq N\cup\{t_i\}$, for all $i$. In both
directions below, every point keeps its cost, so this relaxation and \LPrel{} have the
same optimum value.

\emph{From flows to strolls.} Let $(x,z)$ be feasible, and for each $i$ let
$x^i_{uv}=x_{i,(u,v)}+x_{i,(v,u)}$, $y^i_v=z_{iv}$ for $v\in N$,
$y^i_{s_i}=y^i_{t_i}=\frac12$, and $y^i_v=0$ for the other terminals. The degree
constraints hold, because every $v\in N$ has in-flow and out-flow $z_{iv}$, the vertex $s_i$
has out-flow~$1$ and no in-flow, and $t_i$ has in-flow~$1$ and no out-flow. For $U\subseteq V$
and $U_i=U\cap V(D_i)$, we have $x^i(\delta(U))=x_i(\din(U_i))+x_i(\dout(U_i))$, and flow
conservation gives $x_i(\din(U_i))-x_i(\dout(U_i))=\one[t_i\in U]-\one[s_i\in U]$. Hence
$x^i(\delta(U))\ge1$ if $s_i\in U\not\ni t_i$, and $x^i(\delta(U))=2x_i(\din(U_i))\ge2y^i_v$ if
$v\in U\subseteq V\setminus\{s_i,t_i\}$ and $v\in N$. So $(x^i,y^i)_i$ is feasible for
\LPrel{}.

\emph{From strolls to flows.} Let $(x^i,y^i)_i$ be feasible for \LPrel{}, fix $i$, and write
$x=x^i$ and $y=y^i$. For the terminals $v\notin\{s_i,t_i\}$, we have $y_v=0$ and hence
$x(\delta(v))=0$, so $x$ vanishes outside $V(D_i)$. Let $b_{s_i}=0$, $b_{t_i}=1$, and $b_v=y_v$ for
$v\in N$. For $U\subseteq V(D_i)$, let $E_U$ be the set of edges with both endpoints in $U$;
the degree constraints give $x(E_U)=\sum_{v\in U}y_v-\frac12x(\delta(U))$, and $\sum_{v\in U}(y_v-b_v)=\frac12(\one[s_i\in U]-\one[t_i\in U])$, which is
at most $\frac12x(\delta(U))$ by the cut constraint for the sets that contain $s_i$ but not
$t_i$. Hence $x(E_U)\le\sum_{v\in U}b_v$, with equality for $U=V(D_i)$. By the max-flow
min-cut theorem, applied to the network in which a source sends $x_e$ units to each edge
$e$, each edge forwards them to its endpoints, and each vertex $v$ sends $b_v$ units to a
sink, we can therefore split every $x_{uv}$ into $x_{i,(u,v)}+x_{i,(v,u)}$ such that every
vertex $v\in V(D_i)$ has in-flow $b_v$ and out-flow $x(\delta(v))-b_v$. These are $0$ and $1$ at
$s_i$, $1$ and $0$ at $t_i$, and $y_v$ and $y_v$ at $v\in N$, so $x_i$ is a flow in $D_i$
that satisfies the flow constraints with $z_{iv}=y_v$. For $v\in N$ and
$v\in U\subseteq N\cup\{t_i\}$, flow conservation gives
$x_i(\din(U))=\frac12\bigl(x(\delta(U))+\one[t_i\in U]\bigr)$, which is at least $y_v$: if
$t_i\notin U$, by the cut constraint $x(\delta(U))\ge2y_v$, and if $t_i\in U$, by the cut
constraint $x(\delta(U))\ge1$ for $V\setminus U\ni s_i$ and by $y_v\le1$. So $(x,z)$ is
feasible.

\section{\texorpdfstring{Completion: proof of Lemma~\ref{lem:completion}}{Completion}}\label{app:completion}

\begin{restate}{Lemma~\ref{lem:completion}}
\LemCompletion
\end{restate}

\begin{proof}
Every vertex has even degree in the multigraph $K\uplus J$, because $J$ is an
$\odd(K)$-join, and every component of $K\uplus J$ contains a vertex of $V(Q)$, because
every component of $K$ does. For every component $C$ with at least one edge, choose a
vertex $v_C\in V(Q)\cap V(C)$ and a trunk $Q_i$ that contains $v_C$, and insert a closed
Euler tour of $C$, starting and ending at $v_C$, into $Q_i$ at an occurrence of $v_C$.
This gives, for every $i$, a walk $W_i$ from $s_i$ to $t_i$; together the walks visit
every vertex, since $K$ spans $V$, and their total cost is $c(Q)+c(K)+c(J)$.

For Multi-Path TSP, remove from every $W_i$ all inner occurrences of terminals, and
then all but one occurrence of every vertex of $N$ among all walks. Each removal
replaces two consecutive edges $xy,yz$ by the edge $xz$, which does not increase the
cost by the triangle inequality. Every terminal remains visited as an endpoint of its
own walk, so the result is a family of $s_i$--$t_i$ paths that visit all vertices.

For Ordered TSP, concatenate $W_1,\dots,W_k$ to a closed walk.
It passes through $d_1,d_2,\dots,d_k$, in this cyclic order, at the points where
consecutive walks are joined. Remove all other occurrences of ordered vertices and all
but one occurrence of every vertex of $N$; as before, this does not increase the cost,
and the result is a Hamiltonian cycle that visits $d_1,\dots,d_k$ in this cyclic order.
All steps take polynomial time.\qed
\end{proof}

\section{\texorpdfstring{Forests and derandomization: proofs of Lemma~\ref{lem:kruskal} and Theorems~\ref{thm:mp} and~\ref{thm:ot}}{Forests and derandomization}}\label{app:tools}

\begin{restate}{Lemma~\ref{lem:kruskal}}
\LemKruskal
\end{restate}

\begin{proof}
Run Kruskal's algorithm on $G$ with weights $w$, skipping edges that
would join two components that both meet $A$. By the correspondence
with spanning trees of $G/A$, the resulting forest $F$ has cost $f_w(A)$.

Let $n(t)$ be the number of rootless components of $G_t$.
After Kruskal has processed all edges of weight at most $t$, the current
forest has the same rootless components as $G_t$, and each further accepted
edge will reduce their number by one. Thus
exactly $n(t)$ edges of $F$ have weight greater than $t$. Integrating
over $t$ gives
\[
  f_w(A)=\sum_{e\in F} w_e
        =\int_0^\infty |\{e\in F:w_e>t\}|\,dt
        =\int_0^\infty n(t)\,dt .
\]For the second claim, take $w=c$ and $A=T$. Every $U\in\Uc_c(t)$
satisfies $X(\delta(U))\ge2$ by Lemma~\ref{lem:decomp}(d).
Every edge leaving $U$ has cost greater than $t$, and the components
in $\Uc_c(t)$ are disjoint. Thus
\[
  |\Uc_c(t)|\le\tfrac12\sum_{U\in\Uc_c(t)}X(\delta(U))
              \le X(\{e:c_e>t\}).
\]
Integrating gives $f_c(T)\le c(X)=\optlp$.
\qed
\end{proof}

\paragraph{Derandomization.} The algorithms use randomness only to draw the trees, and
each candidate costs at most $c(Q)+r\,f_w(Q)$ plus a term that does not depend on the
trees: the doubling candidate with $r=2$ and $w=c$, the joining tour with $r=1$ and $w=c$,
and the reweighted candidate with $r=\frac2{1+\lambda}$ and the reweighted cost $w$,
by~\eqref{eq:combination}. The analyses use only the expectation of $c(Q)+r\,f_w(Q)$. By
Lemma~\ref{lem:kruskal}, this expectation is $\E[c(Q)]$ plus a weighted sum of the
probabilities of the events $Q\cap U=\emptyset$ for polynomially many sets $U$, so the
method of conditional expectations applies.

\begin{lemma}\label{lem:derand}
Let $w\in\mathbb{Q}^E_{\ge0}$ and a rational $r\ge0$ be given. In polynomial time we can
compute, for every $i$, a path $\hat Q_i$ in the support of the trunk $Q_i$, such that
$\hat Q=\biguplus_i\hat Q_i$ satisfies
$c(\hat Q)+r\,f_w(\hat Q)\le\E\bigl[c(Q)+r\,f_w(Q)\bigr]$.
\end{lemma}

\begin{proof}
Let $\mathcal H$ be the family of all sets that belong to $\Uc_w(t)$ for some $t\ge0$,
and for $U\in\mathcal H$ let $a_U$ be the length of the set of thresholds $t$ with
$U\in\Uc_w(t)$; this set is an interval, because the graphs $G_t$ grow with $t$, and it
is bounded because the complete graph is connected and contains terminals. The family
$\mathcal H$ is laminar, so $|\mathcal H|\le2|V|$, and $\mathcal H$ and the numbers
$a_U$ can be computed by sorting the edges by weight. By Lemma~\ref{lem:kruskal}, every
$A$ with $T\subseteq A\subseteq V$ satisfies
$f_w(A)=\sum_{U\in\mathcal H}a_U\one[A\cap U=\emptyset]$. Since $T\subseteq V(Q)$, this
applies to $A=V(Q)$, and $Q\cap U=\emptyset$ holds exactly when every trunk avoids $U$.
Therefore, the conditional expectation of $c(Q)+r\,f_w(Q)$ given $Q_i=\hat Q_i$ for
$i\le j$,
\[
 \sum_{i\le j}c(\hat Q_i)+\sum_{i>j}\E[c(Q_i)]
 +r\sum_{\substack{U\in\mathcal H\\ U\cap\hat Q_i=\emptyset\ \forall i\le j}}
 a_U\prod_{i>j}\Prob\bigl[Q_i\cap U=\emptyset\bigr],
\]
can be computed in polynomial time, because the distributions of Lemma~\ref{lem:decomp}
have polynomial support and rational probabilities by its full statement in
Appendix~\ref{app:lp}; the products involve at most $k$
rational probabilities of polynomial encoding length. This conditional expectation is the
average, weighted by the probabilities of the possible values of $Q_{j+1}$, of the
conditional expectations given in addition the value of $Q_{j+1}$. Choosing as $\hat
Q_{j+1}$ a value that minimizes the latter therefore does not increase it. After $k$ such
choices, the paths $\hat Q_1,\dots,\hat Q_k$ satisfy the claimed inequality.\qed
\end{proof}

\begin{proof}[of Theorems~\ref{thm:mp} and~\ref{thm:ot}]
In each candidate, use the paths $\hat Q_i$ of Lemma~\ref{lem:derand}, for its $w$ and $r$,
as trunks instead of drawing the trees; as every $\hat Q_i$ lies in the support of $Q_i$ and
the trees are independent, $(\hat Q_1,\dots,\hat Q_k)$ is an outcome of the trunks, so
Lemma~\ref{lem:completion} applies. Then each candidate costs at most the bound that
Lemma~\ref{lem:doubling}, the proof of Lemma~\ref{lem:ot} or Lemma~\ref{lem:master} gives
for its expected cost, so the proofs of Lemmas~\ref{lem:ot} and~\ref{lem:mp} apply to the
cheaper one; all steps run in polynomial time.\qed
\end{proof}

\section{\texorpdfstring{The growth process: proofs of Lemmas~\ref{lem:source} and~\ref{lem:parity}}{The growth process}}\label{app:growth}

\emph{Running time.} The process adds at most $|N|$ edges, and the load of every edge
reaches its cost at most once. By~(ii) and~(iii) below, every event time $t$ is
$c_{uv}$, or $(1+\lambda)c_{uv}-\lambda h(u)$ for an edge $uv$ and a label $h(u)<t$ (if
$h(u)=t$, the second expression equals $c_{uv}$). Hence every label arises from $0$ by
at most $|N|$ such steps, so for rational $\lambda$ all numbers are rational of polynomial
encoding length, and the next event can be found by comparing $O(|V|^2)$ such numbers.

\begin{figure}[t]
\centering
\begin{tikzpicture}[x=1.05cm,y=0.75cm,font=\small]
  % axes
  \draw[->] (0,0) -- (6.4,0) node[below left] {time $t$};
  \draw[->] (0,0) -- (0,4.4) node[left] {load};
  % cost level and label level
  \draw[dashed,black!50] (0,3.5) -- (6.0,3.5);
  \node[left] at (0,3.5) {$c_{uv}$};
  \draw[dashed,black!50] (0,2) -- (2,2);
  \node[left] at (0,2) {$h(u)$};
  % rate-1 continuation (both endpoints unrooted)
  \draw[dotted,thick,black!55] (2,2) -- (3.5,3.5);
  \draw[dotted,black!45] (3.5,3.5) -- (3.5,0) node[below,text=black] {$c_{uv}$};
  % the load
  \draw[very thick] (0,0) -- node[above left,pos=0.55] {rate $1$} (2,2)
        -- node[below right,pos=0.45] {rate $\frac1{1+\lambda}$} (4.25,3.5);
  \fill (4.25,3.5) circle (1.6pt);
  % time marks
  \draw[dashed,black!50] (2,2) -- (2,0) node[below,text=black] {$h(u)$};
  \draw[dashed,black!50] (4.25,3.5) -- (4.25,0) node[below,text=black] {$w_{uv}$};
  % surcharge
  \draw[decorate,decoration={brace,amplitude=4pt}] (5.2,3.5) -- (5.2,2)
        node[midway,right=5pt] {$\theta_{uv}=c_{uv}-h(u)$};
  \draw[dashed,black!35] (2,2) -- (5.2,2);
\end{tikzpicture}
\caption{The load of an edge $uv$ with $h(u)<h(v)$ as a function of the time $t$,
drawn for $\lambda=\frac12$: it grows at rate $1$ while $u$ and $v$ lie in distinct
rootless components, and at rate $\frac1{1+\lambda}$ after $u$ is rooted. If it reaches
$c_{uv}$ before $v$ is rooted, it does so at time
$h(u)+(1+\lambda)(c_{uv}-h(u))=c_{uv}+\lambda\theta_{uv}=w_{uv}$; at rate $1$ throughout
(dotted), it would reach $c_{uv}$ at time $c_{uv}$.}
\label{fig:load}
\end{figure}
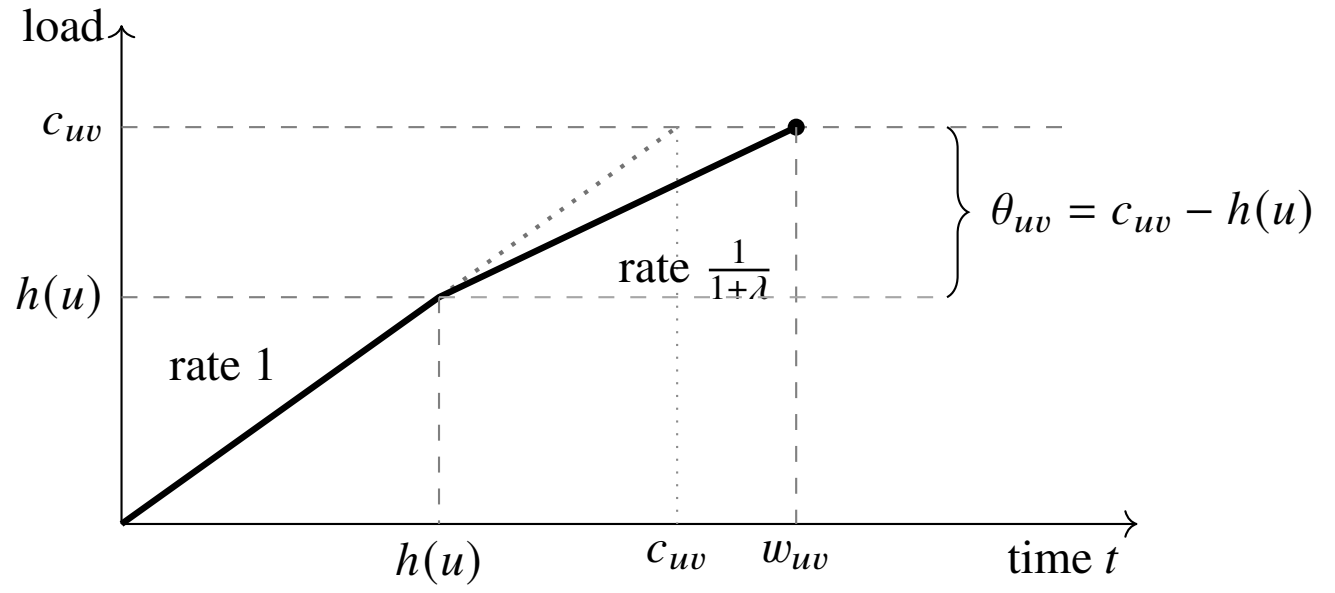

We use three properties of the process. (i)~No load exceeds the cost of its edge: the
load of an edge grows only while the edge joins two rootless components or a rootless
component and a rooted vertex, and when the load reaches the cost, the edge is added
to $F$ or its endpoints have been connected or both rooted; in either case the load
stops growing. (ii)~If $u$ and $v$ lie in distinct rootless components throughout the
interval $[0,t)$, then $\zeta_{uv}(t)=t$, because $uv$ lies in the
boundaries of both components and in no other boundary. (iii)~If $u$ is rooted at time
$h(u)$, and $u$ and $v$ lie in distinct components during $[0,h(u))$ and $v$ is not
rooted before time $t\ge h(u)$, then
$\zeta_{uv}(t)=h(u)+\frac{t-h(u)}{1+\lambda}$: the load is $h(u)$ at time $h(u)$ by~(ii), and after time
$h(u)$ the edge lies in the boundary of the rootless component of $v$ and in the
boundary of $U_t$, which gives rate $\frac12+\frac{1-\lambda}{2(1+\lambda)}=\frac1{1+\lambda}$. An edge that
joins two rootless components when its load reaches its cost, one of which is rooted
by a tied edge processed earlier, is added as an edge between a rooted vertex $u$ and a
rootless component, with $h(u)=t$; (iii) covers this case.
Figure~\ref{fig:load} illustrates~(ii) and~(iii).

\begin{restate}{Lemma~\ref{lem:source}}
\LemSource
\end{restate}

\begin{proof}
\emph{The weights of the forest edges.} Let $e=uv$ be added to $F$ at time $t$. If $e$
joins two rootless components, then $c_e=t$ by~(ii), and both endpoints are rooted at
time $t$ or later, so $\min\{h(u),h(v)\}\ge c_e$, $\theta_e=0$ and $w_e=t$. If $e$
joins a rooted vertex $u$ to a rootless component, then $u$ and $v$ were in distinct
components before time $h(u)$, and~(iii) gives $c_e=h(u)+\frac{t-h(u)}{1+\lambda}$,
so $c_e\ge h(u)$; as $h(v)=t\ge h(u)$, we get $\theta_e=c_e-h(u)$ and
$w_e=(1+\lambda)c_e-\lambda h(u)=t$.

\emph{Minimality.} Let $e=uv$ be an edge not in $F$ whose endpoints are not both
terminals, and let $\tau$ be the first time at which $u$ and $v$ lie in a common
component of $F$ or are both rooted. The path between $u$ and $v$ in the tree obtained
from $F$ by identifying $T$ consists of edges that were added at times at most $\tau$,
so these edges have $w$-cost at most $\tau$. We show $w_e\ge\tau$; then the cycle
property of minimum spanning trees, applied after identifying $T$, shows that $F$ is a
minimum $w$-cost $T$-rooted forest. Let $h(u)\le h(v)$. Before time $\tau$, the
vertices $u$ and $v$ lie in distinct components. If $h(u)\ge\tau$, then by~(ii) the
load of $e$ at time $\tau$ is $\tau$, so $c_e\ge\tau$ by~(i), and $w_e\ge c_e\ge\tau$.
Otherwise $h(u)<\tau$ and $v$ is not rooted before time $\tau$, so~(iii) and~(i) give
$h(u)+\frac{\tau-h(u)}{1+\lambda}\le c_e$, that is,
$\tau\le(1+\lambda)c_e-\lambda h(u)\le c_e+\lambda(c_e-h(u))^+=w_e$, since
$\min\{h(u),h(v)\}=h(u)$.

\emph{The cost of the forest.} Every edge added to $F$ reduces the number of rootless
components by one, and at the end there are none. Hence, for every $t$, the number of
edges added after time $t$ is $|\Ac_t|$, and
\[w(F)=\sum_{e\in F}w_e=\int_0^\infty|\{e\in F:w_e>t\}|\,dt=\int_0^\infty|\Ac_t|\,dt.\]

\emph{The dual budget.} 
For $A\subseteq N$, write $M(A)=\sum_{i=1}^k \max_{v\in A} y^i_v$,
with empty maxima taken to be zero. By the definition of the fractional
sum of peaks, $H=\int_0^\infty M(U_t)\,dt$.

For every nonempty $A\subseteq N$, the stroll cut constraints give
$x^i(\delta(A))\ge 2\max_{v\in A}y^i_v$. Summing over $i$ yields
$X(\delta(A))\ge 2M(A)$. Moreover, for any $v\in A$ we have
$M(A)\ge M(\{v\})=\sum_i y^i_v=1$.
By~(i), $\zeta_e\le c_e$ for every edge $e$. Integrating the growth
rates against $X$ therefore gives
\[
 \begin{aligned}
 \optlp\ge\sum_eX_e\zeta_e
 &=\int_0^\infty\left(\frac12\sum_{A\in\Ac_t}X(\delta(A))
       +\frac{1-\lambda}{2(1+\lambda)}X(\delta(U_t))\right)dt\\
 &\ge f_w(T)+\frac{1-\lambda}{1+\lambda}\int_0^\infty M(U_t)\,dt
 =f_w(T)+\frac{1-\lambda}{1+\lambda}H.
 \end{aligned}
\]
This also covers $N=\emptyset$.\qed
\end{proof}

To prove the parity bound, we use the following submodular packing--covering
lemma. The covering inequalities ensure a valid fractional join; the packing
inequalities control its cost at every level of the labels.

\begin{lemma}[Packing and covering]\label{lem:packing-lp}
Let $(U,F)$ be a finite loopless multigraph and let $M:2^U\to\R_{\ge0}$ be
monotone and submodular, with $M(\emptyset)=0$ and $M(\{v\})\ge1$ for every $v\in U$.
There is $r\in[0,1]^F$ such that
\[
 r(F[A])\le M(A)-1\qquad(\emptyset\ne A\subseteq U),
\]
and, for every $S\subseteq U$ with $\delta_F(S)\ne\emptyset$,
\[
 r(\delta_F(S))\ge M(U)+1-M(S)-M(U\setminus S).
\]
Here $F[A]$ is the multiset of edge occurrences with both endpoints in $A$, and
$\delta_F(S)$ consists of those with exactly one endpoint in $S$.
\end{lemma}

\begin{proof}
Choose a vector maximizing $\sum_{a\in F}r_a$ subject to the packing inequalities
and $0\le r\le1$. Zero is feasible and the feasible region is compact, so such a
vector exists and no coordinate can be increased while preserving feasibility.
Fix a nonempty cut. If a crossing edge has weight one, its weight suffices because
submodularity gives $M(U)\le M(S)+M(U\setminus S)$.
Otherwise choose a crossing edge. Its coordinate is below one, so a tight packing
set $A$ must contain both endpoints. Put $A_1=A\cap S$ and $A_2=A\setminus S$;
both are nonempty. Nonnegativity and the packing inequalities give
\[
 \begin{aligned}
 r(\delta_F(S))
 &\ge r(F[A])-r(F[A_1])-r(F[A_2])\\
 &\ge M(A)+1-M(A_1)-M(A_2).
 \end{aligned}
\]
Submodularity applied to $(S,A)$ and $(U\setminus S,S\cup A)$ gives, after addition,
\[
 M(S)+M(U\setminus S)+M(A)\ge M(U)+M(A_1)+M(A_2).
\]
The covering inequality follows.\qed
\end{proof}

\begin{restate}{Lemma~\ref{lem:parity}}
\LemParity
\end{restate}

\begin{proof}
The fractional join will use half of the strolls, half of the direct pair edges,
and selected fractions of the edges of $K$. Partition $K$ into $K_T$, the edge
occurrences incident to $T$, and $K_N$, those with both endpoints in $N$.
Each function $A\mapsto\max_{v\in A}y^i_v$ is monotone and submodular,
so their sum $M$ is as well. Since $M(\emptyset)=0$ and $M(\{v\})=1$
for every $v\in N$, Lemma~\ref{lem:packing-lp} applies to $(N,K_N)$.
Let $D$ be the multiset of direct edges $s_it_i$, so $c(D)=\Delta$. Consider
\[
 z=\tfrac12(X+\chi^D)+\chi^{K_T}+\sum_{a\in K_N}r_a\chi^{\{a\}}.
\]

\emph{The join inequalities.} Fix $B\subseteq V$ with $|B\cap\odd(K)|$ odd;
then $|\delta_K(B)|$ is odd. If an edge of $K_T$ crosses $B$, its unit contribution
already gives $z(\delta(B))\ge1$. Otherwise put $S=B\cap N$. Its cut in $K_N$
is nonempty and is exactly $\delta_K(B)$. For each pair $i$, write
\[
 a_i=\max_{v\in S}y^i_v,\qquad b_i=\max_{v\in N\setminus S}y^i_v,
\]
with empty maxima zero. The packing lemma supplies
\[
 r(\delta_{K_N}(S))\ge M(N)+1-M(S)-M(N\setminus S)
                 =1-\sum_i\min\{a_i,b_i\}.
\]
If both endpoints of pair $i$ lie outside $B$, its stroll contributes at least
$a_i$ to $X(\delta(B))/2$; if both lie inside, it contributes at least $b_i$,
by applying the same cut constraints to the complement.
If the pair is separated, its stroll and its direct edge together contribute at
least one to $(X+\chi^D)(\delta(B))/2$. This is at least $\min\{a_i,b_i\}$,
since $0\le y^i_v\le1$ follows from $\sum_i y^i_v=1$ on $N$.
Thus $z(\delta(B))\ge1$ in every case, and the join theorem of Edmonds and
Johnson~\cite{EJ} gives $m(\odd(K))\le c(z)$.

\emph{The cost.} If $a\in K_T$, a terminal endpoint has label zero and
$c_a=\theta_a$. An edge $uv\in K_N$ satisfies
$c_{uv}\le\min\{h(u),h(v)\}+\theta_{uv}$. The time for which both endpoints
belong to $U_t$ is $\min\{h(u),h(v)\}$. Hence, using $r\le1$ and the packing
inequalities whenever $U_t\ne\emptyset$,
\[
 \begin{aligned}
 \sum_{a\in K_N}r_ac_a
 &\le\theta(K_N)+\int_0^\infty r(K_N[U_t])\,dt\\
 &\le\theta(K_N)+H.
 \end{aligned}
\]
For empty $U_t$ the integrand is zero. Adding the other terms of $c(z)$ proves
the bound. The vector $r$ is a certificate only; the algorithm computes a minimum
join and never needs to construct $r$.\qed
\end{proof}

\section{The numerical constants}\label{app:constants}

\emph{Bounds on $e$.} Since $e=\sum_{j\ge0}1/j!$ and
$\sum_{j\ge8}1/j!<\frac1{8!}\sum_{i\ge0}9^{-i}<\frac2{8!}$,
\[
 2.71825<\tfrac{685}{252}=\sum_{j=0}^{7}\tfrac1{j!}<\tfrac{109601}{40320}=\sum_{j=0}^{8}\tfrac1{j!}<e
 <\tfrac{685}{252}+\tfrac2{8!}<2.71831 .
\]
Hence $0.3678<\frac1e<0.3679$.

\emph{Decimal bounds.} The functions $e\mapsto\frac{5e-2}{3e-2}$ and
$e\mapsto1+\frac{e}{2(e-1)}$ are decreasing, with derivatives $-\frac4{(3e-2)^2}$ and
$-\frac1{2(e-1)^2}$. Using $e>\frac{109601}{40320}$ in the first and $e>\frac{685}{252}$ in
the second gives
\[
 \frac{5e-2}{3e-2}<\frac{467365}{248163}<1.8833,\qquad
 1+\frac{e}{2(e-1)}<1+\frac{685}{866}<1.7910 .
\]
The first bound has slack only about $1.5\cdot10^{-6}$, which is why it uses the
sharper lower bound on $e$.
For comparison, $1+2e^{-1/2}=2.21306\ldots$ and $\frac32+\frac1e=1.86787\ldots$.

Proposition~\ref{prop:single} describes the guarantee of the reweighted candidate alone.
\begin{proposition}\label{prop:single}
For every rational $\lambda$ with $0<\lambda<\lambda^*=\frac12-\frac1e$, the reweighted
candidate alone satisfies $\E[\Alam]\le\frac{2+\lambda}{1+\lambda}\,\optlp$.
\end{proposition}

\begin{proof}
Since $\frac1e<\frac12$, we have $\lambda^*<1-\frac2e$, so Lemma~\ref{lem:master} applies, and
$\kappa_\lambda=\frac{2(\lambda-\lambda^*)}{1+\lambda}\le0$. The function
$g(\lambda)=\frac{1+\lambda}e-3\lambda-\lambda^2$ is decreasing, and
$g(\lambda^*)=\frac{11}{2e}-\frac2{e^2}-\frac74$; as a function of $\frac1e$, this increases on
$[0,1]$ and is positive at $\frac1e=0.3678$ (it equals $0.0023\ldots$ there). Hence $g\ge0$ and
$\pi_\lambda=\frac{2g(\lambda)}{1-\lambda^2}\ge0$ on $(0,\lambda^*]$. Lemma~\ref{lem:master} with
$\Delta\ge0$, $f_w(T)\le \optlp$ (Theorem~\ref{thm:reweight}) gives
$\E[\Alam]\le\bigl(\frac{2(1-1/e)}{1-\lambda}+\pi_\lambda\bigr)\optlp+\frac{\lambda(3+\lambda)}{1-\lambda^2}\optlp$.
A direct computation gives, for every $\lambda\in(0,1)$,
\begin{equation}\label{eq:coeff-sum}
 \frac{2(1-\frac1e)}{1-\lambda}+\pi_\lambda=\frac{2(1-2\lambda-\lambda^2)}{1-\lambda^2},\qquad
 \frac{2(1-\frac1e)}{1-\lambda}+\pi_\lambda+\frac{\lambda(3+\lambda)}{1-\lambda^2}=\frac{2+\lambda}{1+\lambda},
\end{equation}
and the first expression is nonnegative because $\lambda\le\lambda^*<\sqrt2-1$. Hence
$\E[\Alam]\le\frac{2+\lambda}{1+\lambda}\optlp$.\qed
\end{proof}

\emph{Lemma~\ref{lem:mp}.} For $\lambda=\frac2{15}\le1-\frac2e$, the coefficients of
Lemma~\ref{lem:master} are $\frac{2(1-1/e)}{1-\lambda}=\frac{30}{13}(1-\frac1e)$,
$\frac{\lambda(3+\lambda)}{1-\lambda^2}=\frac{94}{221}$,
$\kappa_\lambda=\frac1{17}(\frac{30}e-11)$, which is positive because $e<\frac{30}{11}$, and
$\pi_\lambda=\frac1{221}(\frac{510}e-188)$, which is negative because
$e>\frac{510}{188}=2.7127\ldots$. Since $\frac{30}{13}=\frac{510}{221}$, the coefficient of $\optlp$ in
the resulting bound on $\E[\Alam]$ is $\frac{510-510/e+94-188+510/e}{221}=\frac{32}{17}$. The
weights $\frac{30-11e}{6e-4}$ on the bound on $\E[\Adbl]$ and $\frac{17(e-2)}{6e-4}$ on the
bound on $\E[\Alam]$ are positive because $2<e<\frac{30}{11}$, and they sum to one. In their
combination, the coefficient of $\Delta$ is
$\frac{(30-11e)(2-e)}{e(6e-4)}+\frac{(e-2)(30-11e)}{e(6e-4)}=0$, the coefficient of $\optlp$ is
$\frac{2(30-11e)+32(e-2)}{6e-4}=\frac{5e-2}{3e-2}$, and the coefficient of $\optlp-f_c(T)$ is
\begin{align*}
 -\frac2e\cdot\frac{30-11e}{6e-4}+\frac{17(e-2)}{6e-4}\cdot\frac{188e-510}{221e}
 &=\frac{(e-2)(188e-510)-26(30-11e)}{13e(6e-4)}\\
 &=-\frac{2(150e-47e^2-60)}{13e(3e-2)} .
\end{align*}
The quadratic $150e-47e^2-60$ is positive for $0.47<e<2.7226$, hence at~$e$.

\emph{Other parameters.} The proof of Lemma~\ref{lem:mp} works for every rational
$\lambda\in(\lambda^*,1-\frac2e]$ with $(1-\frac2e)\max\{0,-\pi_\lambda\}\le\frac2e\kappa_\lambda$.
Indeed, $\kappa_\lambda>0$ for such $\lambda$. By~\eqref{eq:coeff-sum}, Lemma~\ref{lem:master}
gives $\E[\Alam]\le\frac{2+\lambda}{1+\lambda}\optlp+\kappa_\lambda\Delta$ if $\pi_\lambda\ge0$ (using
$f_w(T)\le \optlp$ and that the first expression in~\eqref{eq:coeff-sum} is nonnegative,
as $\lambda\le1-\frac2e<\sqrt2-1$), and the same bound plus $|\pi_\lambda|(\optlp-f_c(T))$ if
$\pi_\lambda<0$ (using $f_w(T)\ge f_c(T)$). The weights
$\frac{\kappa_\lambda}{1-2/e+\kappa_\lambda}$ on the bound on $\E[\Adbl]$ in the proof of
Lemma~\ref{lem:mp} and $\frac{1-2/e}{1-2/e+\kappa_\lambda}$ on this bound cancel $\Delta$. As
$1-\frac2e+\kappa_\lambda=\frac{\lambda(3-2/e)}{1+\lambda}$ and
$2\kappa_\lambda+(1-\frac2e)\frac{2+\lambda}{1+\lambda}=\frac{\lambda(5-2/e)}{1+\lambda}$, the
combination is $\frac{5e-2}{3e-2}\optlp$ plus
$\frac{(1-2/e)\max\{0,-\pi_\lambda\}-(2/e)\kappa_\lambda}{1-2/e+\kappa_\lambda}(\optlp-f_c(T))\le0$.
The admissible parameters form an interval: $\pi_\lambda$ has the sign of
$\frac{1+\lambda}e-3\lambda-\lambda^2$, which decreases in $\lambda$ and vanishes at
$0.13304\ldots$; and if $\pi_\lambda<0$, multiplying the condition by $\frac{1-\lambda^2}2$
turns it into $\lambda\bigl(\lambda-(\frac{10}e-\frac4{e^2}-3)\bigr)\le0$. Hence the condition
holds exactly for $\lambda\in(\lambda^*,\frac{10}e-\frac4{e^2}-3]=(0.13212\ldots,0.13745\ldots]$.
For $\lambda=\frac2{15}$, the sign conditions hold with small margins
($\kappa_\lambda\approx0.0021$ and $\pi_\lambda\approx-0.0017$).

\emph{The limit of the analysis.} Let $\lambda\in(0,1-\frac2e]$, and put
$f_c(T)=f_w(T)=\optlp$ and $\E[c(Q)]=\Delta=\optlp/(3-\frac2e)$. Then the bound of
Lemma~\ref{lem:doubling} is $2\optlp-(1-\frac2e)\Delta=\optlp+2\Delta=\frac{5e-2}{3e-2}\optlp$, and
by~\eqref{eq:coeff-sum} the bound of Lemma~\ref{lem:master} is
$\frac{2+\lambda}{1+\lambda}\optlp+\kappa_\lambda\Delta=\optlp+2\Delta$, because
$(2+\lambda)(3-\frac2e)+2\lambda-1+\frac2e=(1+\lambda)(5-\frac2e)$.
\paragraph{Interpretation of the limit.} This is a limit of the displayed
inequalities, not a lower bound on the algorithm. No instance with $\optlp>0$
attains the formal balance: $f_w(T)=\optlp$ would imply $H=0$ by
Lemma~\ref{lem:source}(b). Since $M(U_t)\ge1$ whenever $U_t\ne\emptyset$, all
rootless components would have zero lifetime and $f_w(T)=0$.
Whether instances approach it, and the true ratio of the algorithm, remain open.
A better ratio needs a stronger candidate bound, a further relation between these
quantities, or another candidate.

\end{document}